\documentclass[12pt]{revtex4}
\usepackage[dvips]{graphicx}
\usepackage{delarray}
\usepackage{amsmath}
\usepackage{amssymb}
\usepackage{color}
\usepackage{esint}
\usepackage{float}
\usepackage{calc}
\usepackage{ae}

\newcommand{\be}{\begin{equation}} \newcommand{\ee}{\end{equation}}

\newcommand{\vpa}{v_{\|}}
\newcommand{\vpe}{v_{\perp}}

\newcommand{\mylsim}{\mathrel{\mbox{\raisebox{-1mm}{$\stackrel{<}{\sim}$}}}}

\newcommand{\epso}{\varepsilon_0}

\renewcommand{\P}{\mathcal{P}}
\newcommand{\E}{\bar{E}}
\newcommand{\pmax}{p_\text{max}}
\newcommand{\vpvp}{\vpe/\vpa}

\newcommand{\Bekefi}{\ensuremath{\mathcal{P}_{\mbox{\scriptsize cyl}}\, }}
\newcommand{\FullPank}{\ensuremath{\mathcal{P}_{\mbox{\scriptsize full}}\, }}
\newcommand{\PankOne}{\ensuremath{\mathcal{P}_{\mbox{\scriptsize as1}}\, }}
\newcommand{\PankTwo}{\ensuremath{\mathcal{P}_{\mbox{\scriptsize as2}}\, }}

\begin{document}
\global\long\def\dt{\mbox{d}\tau}
\global\long\def\dy{\mbox{d}y}
\global\long\def\v{\mathbf{v}}
\global\long\def\R{\mathbf{R}}
\global\long\def\B{\mathbf{B}}
\global\long\def\ml{\chi}
\global\long\def\zeff{Z_{\text{eff}}}
\global\long\def\vp{v_{\perp}}
\global\long\def\vpa{v_{\parallel}}
\global\long\def\pp{p_{\perp}}
\global\long\def\ppa{p_{\parallel}}
\global\long\def\vpb{\bar{v}_{\perp}}
\global\long\def\vpab{\bar{v}_{\parallel}}
\global\long\def\tone{\hat{\tau}_{1}}
\global\long\def\ttwo{\hat{\tau}_{2}}
\global\long\def\tz{\boldsymbol{\tau}_{0}}
\global\long\def\ab{\bar{\alpha}}
\global\long\def\tb{\bar{\theta}}
\global\long\def\imag{\mathfrak{Im}}
\global\long\def\power{\mathfrak{Im}}
\global\long\def\power#1{\cdot10^{#1}}

\begin{center}
\Large
 
{\bf Synchrotron radiation from a runaway electron distribution in
  tokamaks}\\ ~\\*[0.5cm] \normalsize {A. Stahl$^1$,
  M. Landreman$^{2}$, G. Papp$^{1,3}$, E. Hollmann$^{4}$,
  T. F\"ul\"op$^1$\\ {\it\small $^1$ Department of Applied Physics,
    Nuclear Engineering, Chalmers University of Technology and
    Euratom-VR Association, SE-412 96 G\"oteborg, Sweden\\$^2$ Plasma Science
    and Fusion Center, MIT, Cambridge, MA, 02139, USA\\$^{3}$ Department of Nuclear Techniques, Budapest University of Technology and Economics, Association EURATOM, H-1111 Budapest, Hungary\\$^{4}$
    Center for Energy research, University of California, San Diego, La Jolla, CA, 92093-0417, USA
}}
\end{center}
\begin{abstract}
  The synchrotron radiation emitted by runaway electrons in a fusion
  plasma provides information regarding the particle momenta and
  pitch-angles of the runaway electron population through the strong
  dependence of the synchrotron spectrum on these
  parameters. Information about the runaway density and its spatial
  distribution, as well as the time evolution of the above quantities,
  can also be deduced.  In this paper we present the synchrotron
  radiation spectra for typical avalanching runaway electron
  distributions.  Spectra obtained for a distribution of electrons are
  compared to the emission of mono-energetic electrons with a
  prescribed pitch-angle.  We also examine the effects of magnetic
  field curvature and analyse the sensitivity of the resulting
  spectrum to perturbations to the runaway distribution.  The
  implications for the deduced runaway electron parameters are
  discussed. We compare our calculations to experimental data from
  DIII-D and estimate the maximum observed runaway energy.
\end{abstract}

\maketitle

\section{Introduction}\label{sec:introduction}
Understanding the process of runaway beam formation and loss in
tokamaks is of great importance, due to the potentially severe damage
these electrons may cause in disruptions.  In present
tokamaks, runaway electrons have energies between a few hundred keV
to tens of MeV, and in a next-step device like ITER, they are
projected to reach a maximum energy of up to 100 MeV \cite{pappRMP}. 
Runaway electrons emit synchrotron radiation
\cite{winske,finken,jaspers,yu},
the spectrum of which 
depends on the
velocity-space distribution of the radiating particles. Therefore, the spectrum
can be used to obtain information about the departure of the velocity
distribution from isotropy and about the energy of the particles. The
emitted radiation can also be an energy loss mechanism
\cite{andersson}, although in tokamaks this loss is not appreciable
unless the electrons have very large energies, above 70 MeV
\cite{finken}.

Many theoretical studies of the synchrotron radiation of the energetic
population have been done before, either using approximate
electron distribution functions or assuming straight magnetic
field lines \cite{robinson,moghaddam,kato}. In several studies, the
synchrotron emission from a single particle is used as an
approximation for the entire runaway distribution \cite{jaspers,yu},
using a specific momentum and pitch-angle for the electrons, often
identified as the maximum momentum and pitch-angle of the electrons in
the runaway beam. In the present work we use an electron distribution
function typical of avalanching runaway electron populations in
tokamak disruptions.  As we will show, taking into account the whole
distribution is important, since synchrotron radiation diagnostics
based on single particle emission can give misleading
results. Furthermore, we will illustrate that synchrotron radiation can
be used to detect signs of modification of the electron distribution,
which can occur due to for instance wave-particle interaction. 

The structure of the paper is as follows. In Sec.~\ref{sec:spectrum}
we give several expressions for the radiated synchrotron power
including the effect of field-curvature. We also discuss the
applicability of these expressions in different contexts.
Section~\ref{sec:distribution} is devoted to the analysis of
the synchrotron radiation spectrum from an avalanching runaway
electron distribution. We will describe the parametric dependences on
magnetic field, density, temperature, effective charge and electric
field. In Sec.~\ref{sec:discussion} we discuss the potential use of
synchrotron radiation as a diagnostic. We also present a
comparison between the synchrotron spectrum calculated for the
avalanching runaway electron distribution and an experimentally
measured synchrotron spectrum from DIII-D. Our conclusions will be
summarized in Sec.~\ref{sec:conclusions}.

\section{Synchrotron emission formulas}\label{sec:spectrum}
\begin{figure}
\includegraphics[trim=0.5cm 0.2cm 0.9cm 0.7cm, clip=true, width=0.48\textwidth]{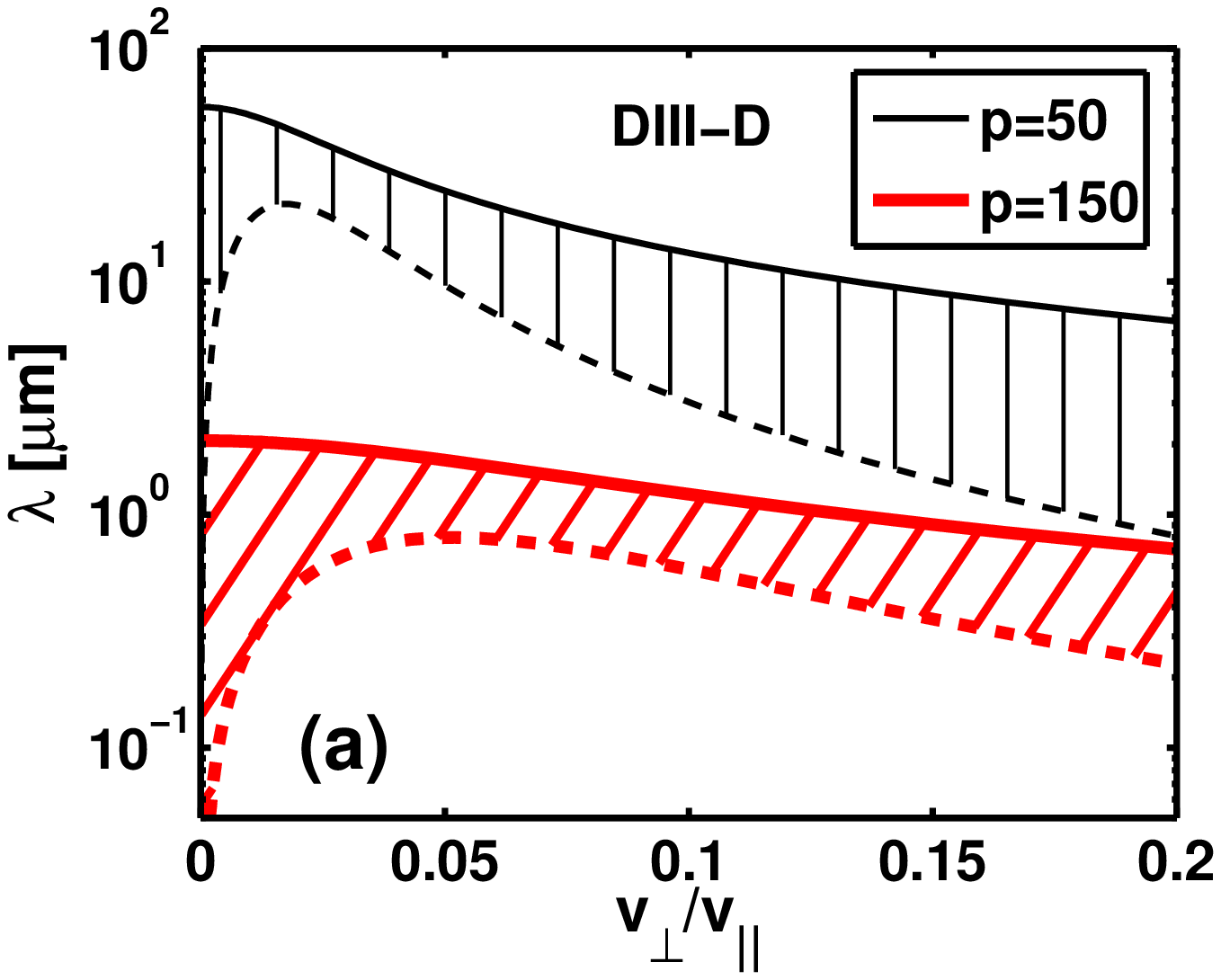}
\includegraphics[trim=0.5cm 0.2cm 0.9cm 0.7cm, clip=true, width=0.48\textwidth]{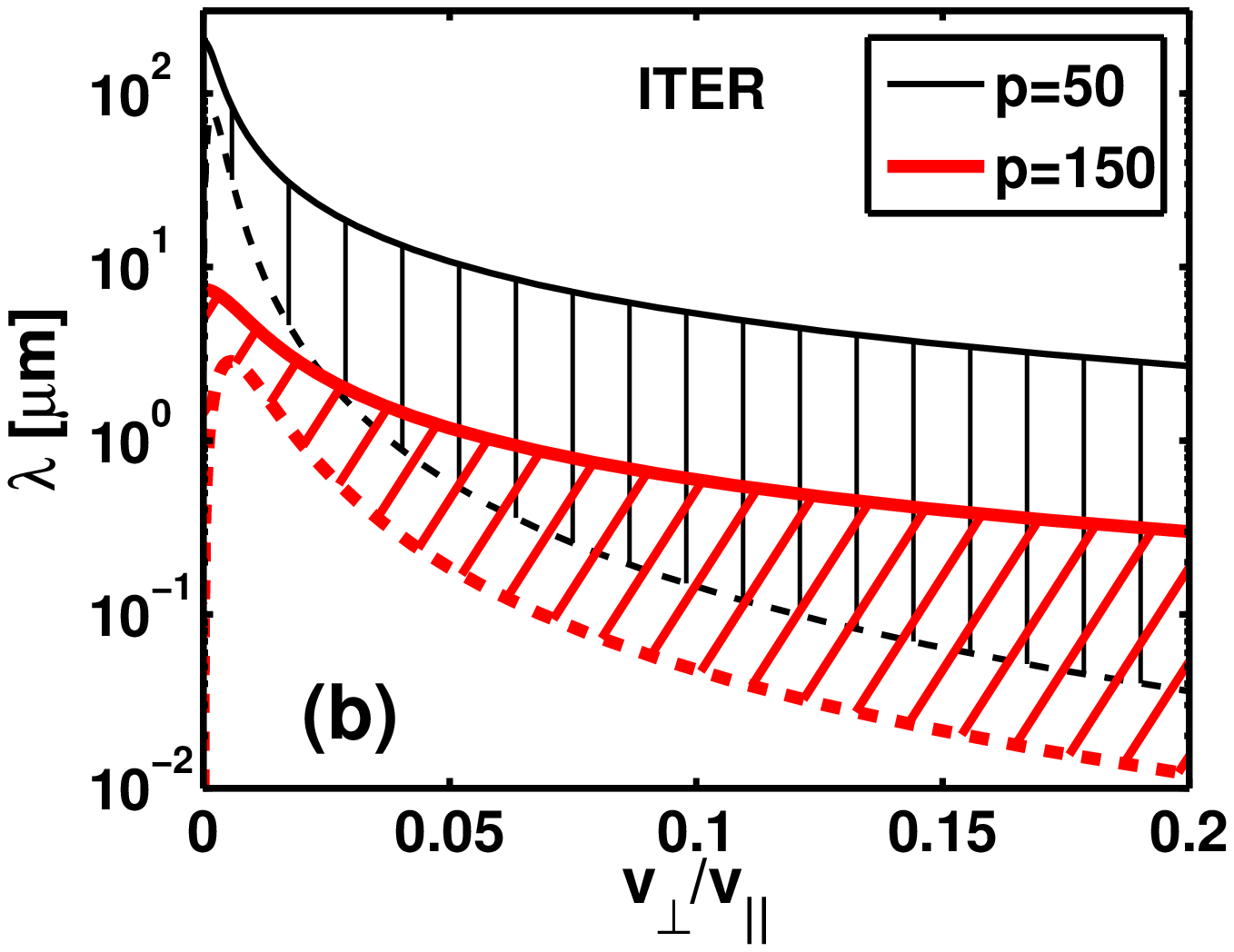}
\caption{(Color online)  Upper and lower bounds on the wavelength $\lambda$
  for which \PankOne is valid. Note the logarithmic scale on the vertical axis.  The parameters used are (a) $B=2.1 \;\rm T$ and $R=1.67 \;\rm m$ and (b) $B=5.3 \;\rm T$ and $R=6 \;\rm m$. }
\label{lambdabounds1}
\end{figure}

The power radiated by an electron with Lorentz factor $\gamma \gg 1$
at wavelength $\lambda$ in the case of straight magnetic field lines
is \cite{bekefi}
\begin{equation}
\Bekefi (\lambda)=\frac{1}{\sqrt{3}}\frac{ce^{2}}{\epsilon_{0}\lambda^{3}\gamma^{2}}\int_{\lambda_{c}/\lambda}^{\infty}K_{5/3}(l)\mbox{d}l\ ,\label{eq:Bekefi}
\end{equation}
where $e$ is the electron charge, $c$ is the speed of light,  $\epsilon_0$ is the vacuum permittivity, \mbox{$
  \lambda_{c}=(4 \pi cm_e\gamma_{\parallel})/(3eB\gamma^{2})$}, \mbox{$\gamma_\|=1/\sqrt{1-v_\|^2/c^2}$},  $m_e$
is the electron rest mass, $B$ is the magnetic field, $_\parallel$ denotes
the component along
the magnetic field and
$K_{\nu}(x)$ is the modified Bessel function of the second kind. The
radiation is emitted in a narrow beam in the parallel direction due to
relativistic effects \cite{bekefi}. In a tokamak, the effects of magnetic field line curvature
and curvature drift have to be taken into
account. This has been done in Ref.~\cite{pankratov}, where the
following expression was obtained
\begin{align}
\FullPank(\lambda)=\frac{ce^{2}}{\epso\lambda^{3}\gamma^{2}} & \left\{ \ \int_{0}^{\infty}\!\frac{\dy}{y}\Bigl(1+2y^{2}\Bigr)\, J_{0}\left(a y^{3}\right)\sin\left(\frac{3}{2}\xi\left(y+\frac{1}{3}y^{3}\right)\right)\right.\ \nonumber \\
 & \quad-\frac{4\eta}{1+\eta^{2}}\int_{0}^{\infty}\!\dy\: y\: J'_{0}\left(a y^{3}\right)\cos\left(\frac{3}{2}\xi\left(y+\frac{1}{3}y^{3}\right)\right)\ -\frac{\pi}{2}\ \Bigg\}\ ,\label{eq:fullPank}
\end{align}
where $a=\xi\eta / (1+\eta^{2})$,
\begin{equation}
\xi =\frac{4\pi}{3}\,\frac{R}{\lambda\gamma^{3}\sqrt{1+\eta^{2}}}\ ,\label{eq:xi}
\end{equation}
\begin{equation}
\eta=\frac{eBR}{\gamma m_e}\,\frac{\vp}{\vpa^{2}}\simeq\frac{\omega_{c}R}{\gamma c}\,\frac{\vp}{\vpa}\ , \label{eq:eta}
\end{equation}
$R$ is the tokamak major radius, $J_\nu (x)$ is the Bessel function and $J_\nu'(x)$ its derivative. 
The integrands in Eq.~\eqref{eq:fullPank} are highly oscillatory and the calculation of synchrotron spectra can become computationally heavy. This motivates examining more approximate formulas which are less complex, especially when considering possible diagnostic applications. In equations (21) and (26) of Ref.~\cite{pankratov}, two limits of
Eq.~(\ref{eq:fullPank}) are given. 
These two limits are obtained by first expanding in $\xi\gg 1$, which can
be translated to a condition for the wavelength $\lambda \ll (4
\pi/3)R/(\gamma^3\sqrt{1+\eta^2})$. 
Then, to obtain the first of the two expressions,
Eq.~(\ref{eq:fullPank}) is also expanded in the smallness of the argument of the
Bessel functions, leading to the condition $\xi\eta\lesssim
1+\eta^{2}$. The resulting approximative formula is
 \begin{equation}
  \PankOne(\lambda)\approx\frac{ce^{2}}{4\epsilon_{0}}\ \sqrt{\frac{2\sqrt{1+\eta^{2}}}{\lambda^{5}R\gamma}}\ e^{-\xi}\left[I_{0}(a)+\frac{4\eta}{1+\eta^{2}}I_{1}(a)\right]\ ,
\label{eq:pankr1}
\end{equation}
where $I_\nu(x)$ is the modified Bessel function.
\PankOne was the expression used to calculate the
synchrotron radiation of an avalanching population of positrons in
Ref.~\cite{fuloppapp} and in fitting of the synchrotron spectrum in
the optical range in DIII-D in Ref.~\cite{yu}.  The two conditions
required for validity of Eq.~(\ref{eq:pankr1}) can be summarized as
$\eta/(1+\eta^2)\mylsim 1/\xi\ll 1$, which leads to a rather narrow
validity range for \PankOne. Figure \ref{lambdabounds1}(a) and (b) show the range of wavelengths for which
\PankOne is valid ($\lambda_l\mylsim
\lambda\ll\lambda_u$) for different runaway momenta in DIII-D-size and
ITER-size tokamaks, respectively. Note that the wavelength should be
much smaller than the solid line(s) in the figure for
\PankOne to be valid. It is clear that for wavelengths in
the 0.1-1 $\mu$m range (as in the measurements described in
Ref.~\cite{yu}), the approximative formula \PankOne is
only valid for particles with large normalized momenta $p=\gamma v/c$, and not necessarily for all values of $\vpe/\vpa$.

\begin{figure}
\includegraphics[width=0.47\textwidth]{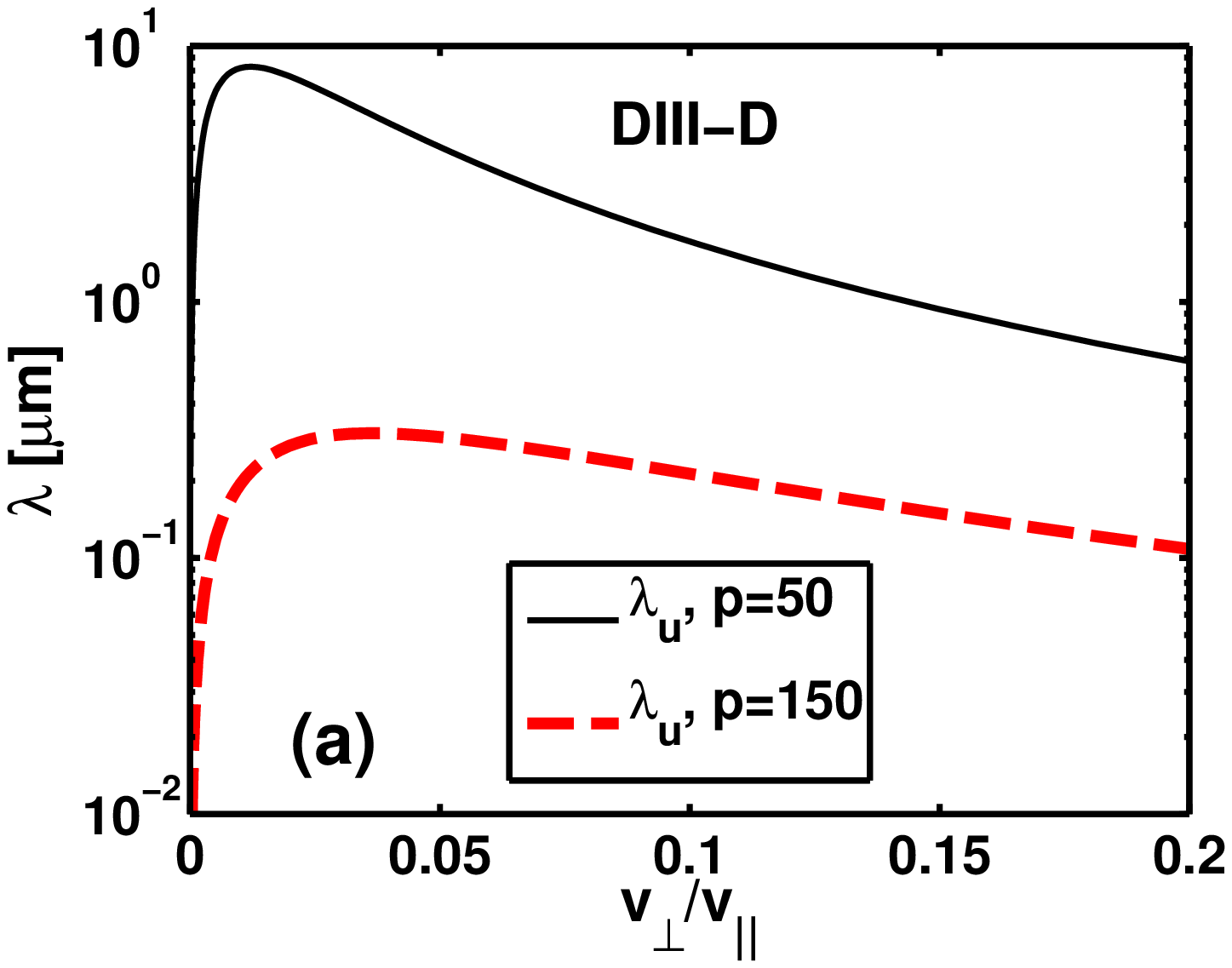}
\includegraphics[width=0.47\textwidth]{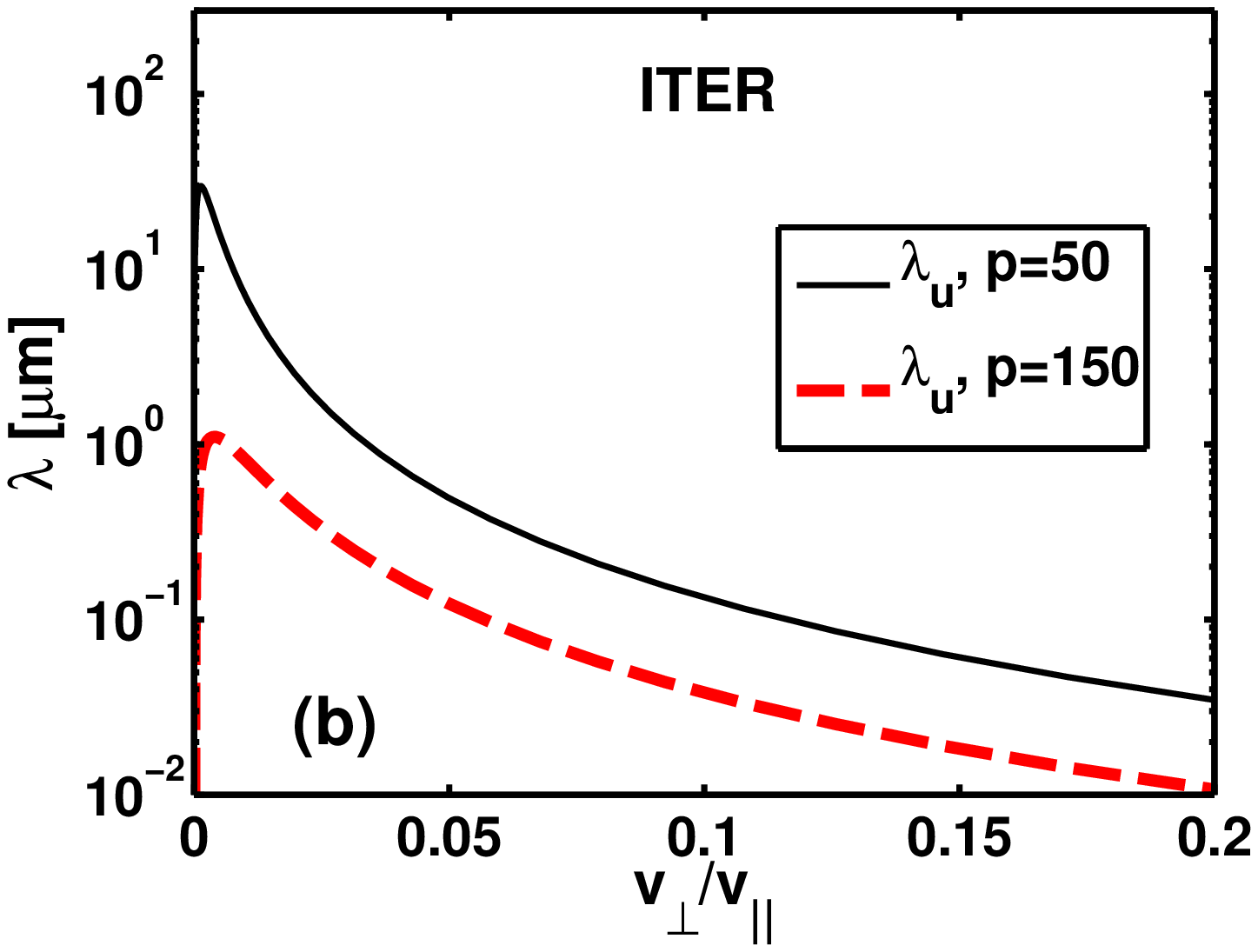}
\caption{(Color online) Upper bounds on the wavelength
  $\lambda$ for which  \PankTwo is valid. Note the logarithmic scale on the vertical axis. The parameters used are (a) $B=2.1 \;\rm T$ and $R=1.67 \;\rm m$ and (b) $B=5.3 \;\rm T$ and $R=6 \;\rm m$. }
\label{lambdabounds2}
\end{figure}

 To obtain the second limit of Eq.~(\ref{eq:fullPank}) (equation (26) in
Ref. \cite{pankratov}), 
\begin{equation} \lambda\ll(4
\pi/3)R\eta/[\gamma^3(1+\eta)^3]
\label{eq:condition26}
\end{equation} 
has to be fulfilled. Equation~\eqref{eq:fullPank} then
simplifies to
\begin{equation}
\PankTwo(\lambda)=\frac{\sqrt{3}}{8\pi}\frac{ce^{2}\gamma}{\epsilon_{0}\lambda^{2}R}\frac{(1+\eta)^{2}}{\sqrt{\eta}}\,\exp\left(-\frac{4\pi}{3}\frac{R}{\lambda\gamma^{3}}\frac{1}{1+\eta}\right)\, .
\label{eq:pankr2}
\end{equation}
The condition in Eq.~\eqref{eq:condition26} is more strict than the
one stemming from $\xi \gg 1$; it is only necessary to fulfill
Eq.~\eqref{eq:condition26} for Eq.~\eqref{eq:pankr2} to be
valid. Figure \ref{lambdabounds2} (a) and (b) show the upper bound for
the wavelength given by Eq.~\eqref{eq:condition26}.  We conclude that
for the visible part of the spectrum, \PankTwo could be a suitable
approximative formula for runaway electron beams with $p < 50$ and
$v_\perp/v_\|<0.1$. In the opposite case, when $p$ and $v_\perp/v_\|$
are large then either the full expression \FullPank, or in
some cases \PankOne, can be used.

\begin{figure}[h]
\begin{center}
\includegraphics[width=0.48\textwidth]{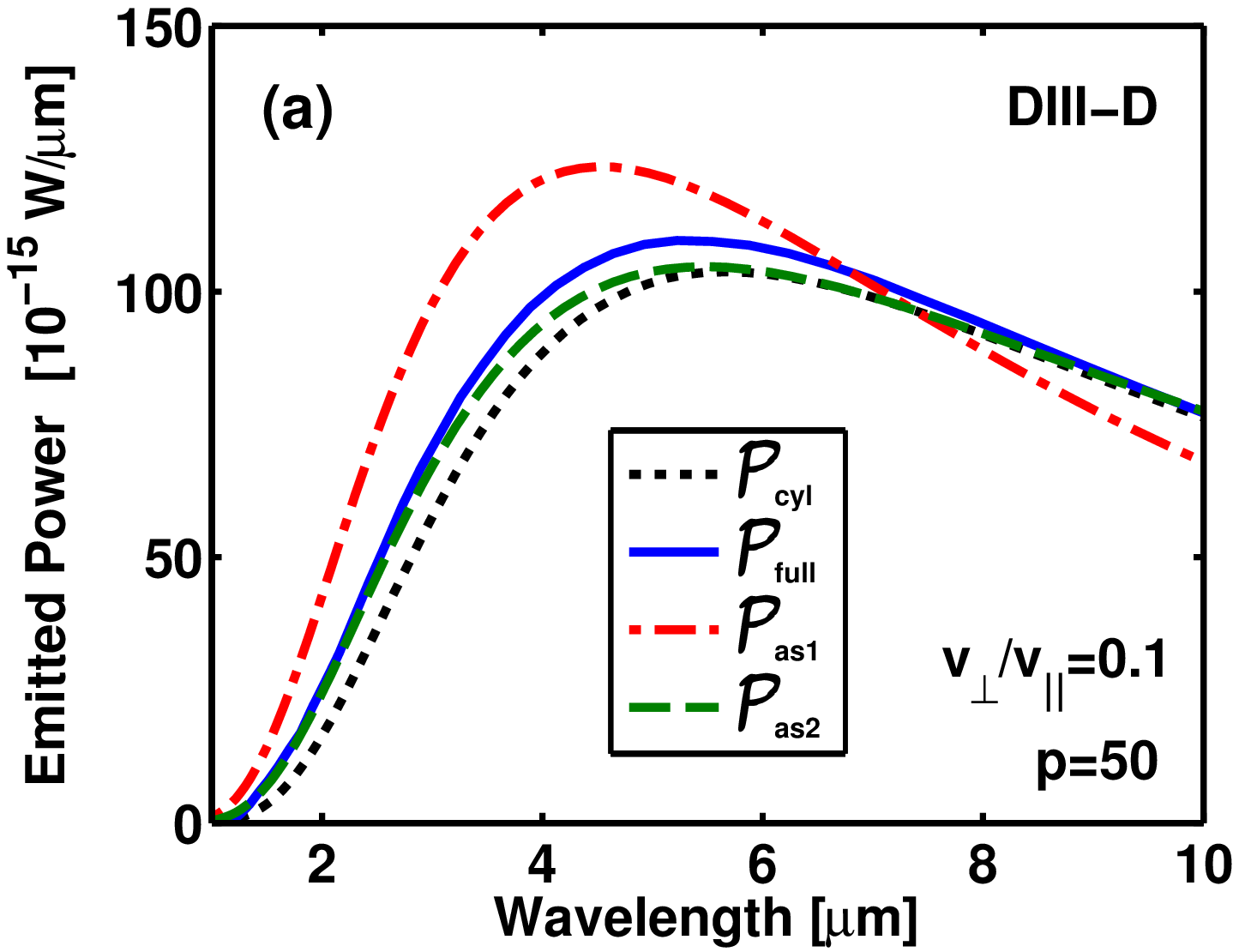}
\includegraphics[width=0.48\textwidth]{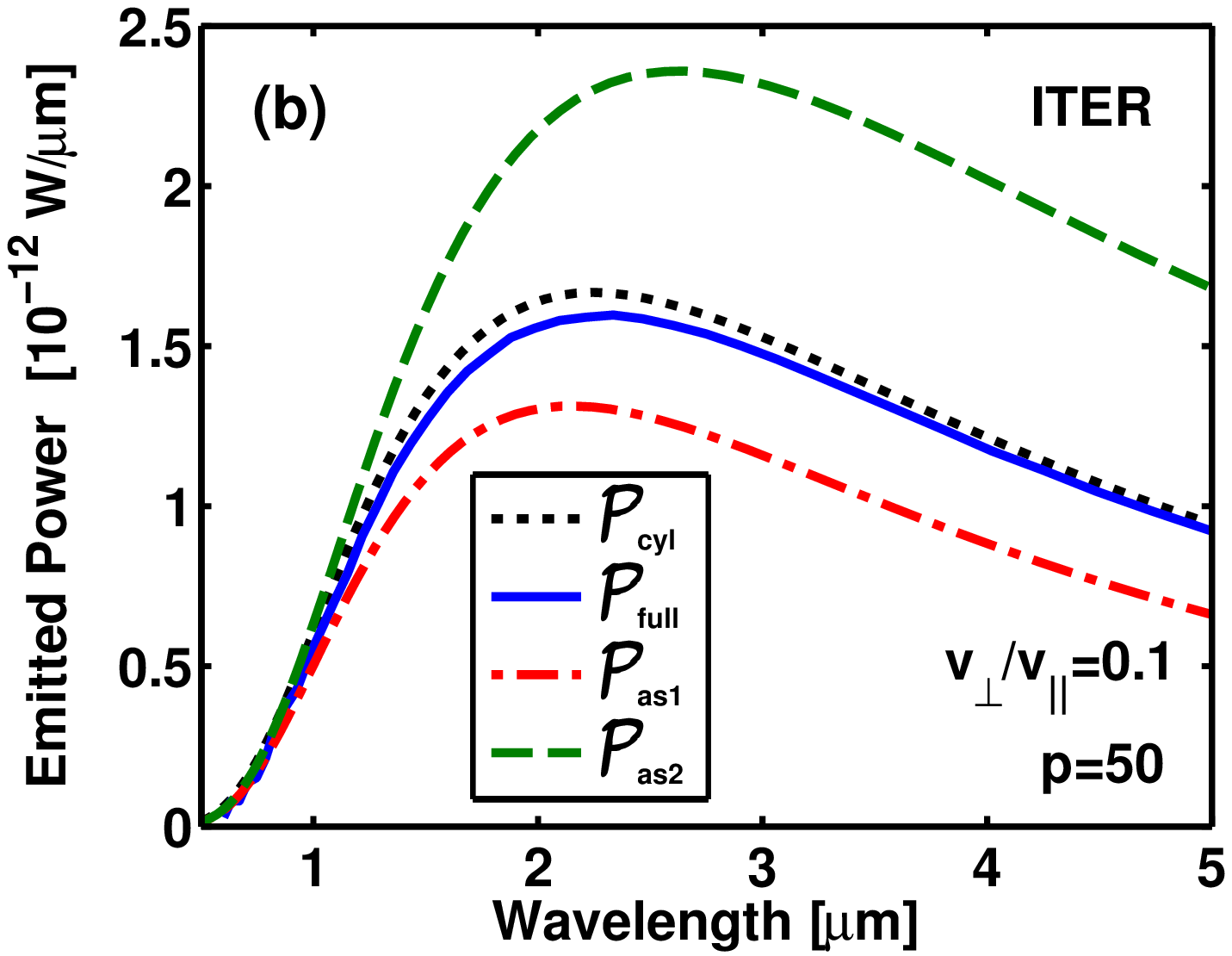}
\includegraphics[width=0.48\textwidth]{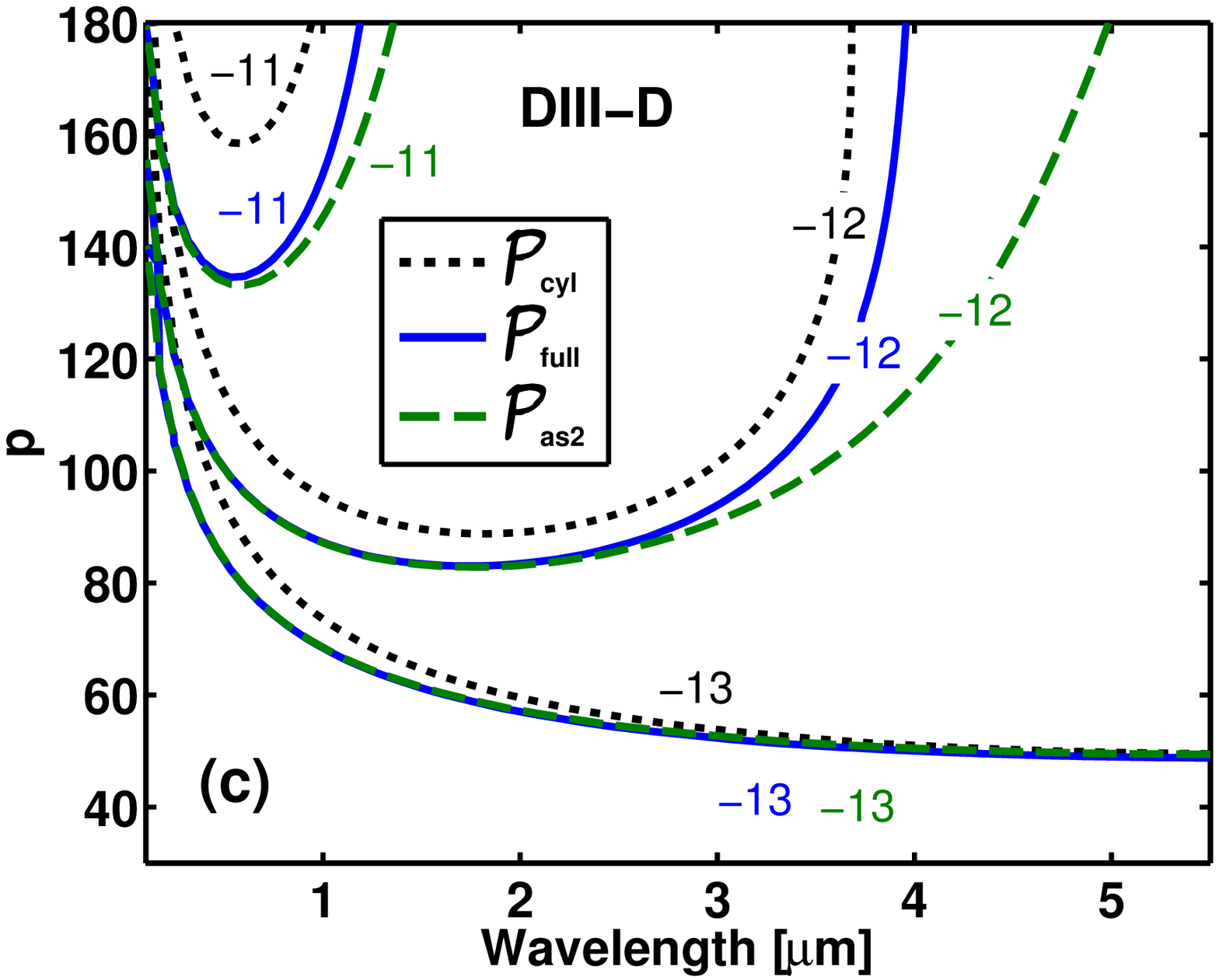}
\includegraphics[width=0.48\textwidth]{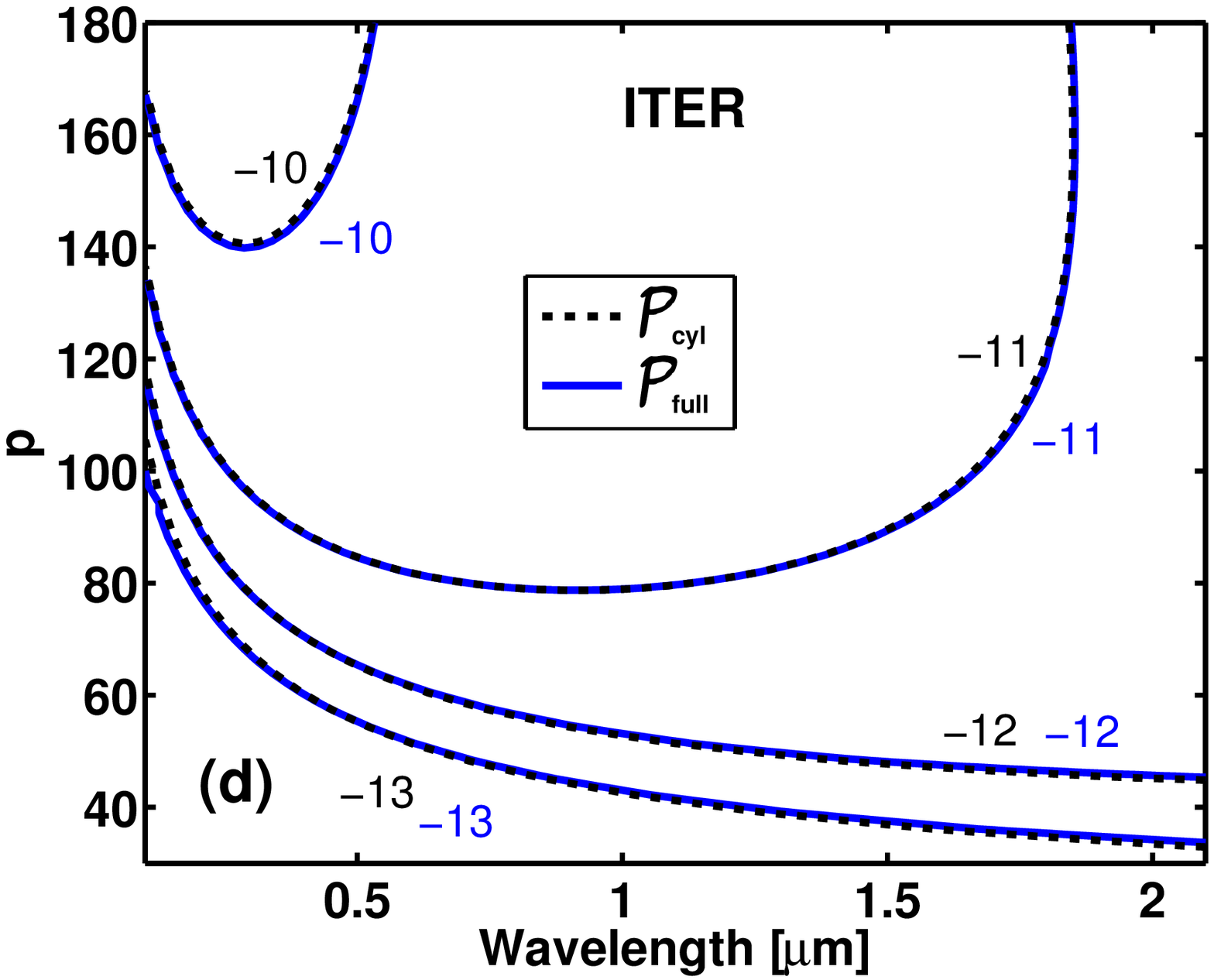}
\caption{(Color online) Single particle synchrotron emission from different emission formulas. (a) and (b) show emitted spectra for particles with
  $\vpe/\vpa=0.1$ and $p=50$ and tokamak parameters corresponding to (a) DIII-D and (b) ITER. The solid (blue) line corresponds to the expression including the field-line curvature, \FullPank. The dotted (black) line is the cylindrical limit, \Bekefi. The dash-dotted (red) and dashed (green) lines correspond to the approximative expressions \PankOne and \PankTwo, respectively. (c) and (d) show contours of $\log_{10}(\mathcal{P}_i(\lambda))$ (with $\mathcal{P}_i$ in units of $W/\mu m$) for various particle momenta and compares (c) \Bekefi, \FullPank and \PankTwo and (b) \Bekefi and \FullPank. }
\label{single_scans}
\end{center}
\end{figure}

In general, the difference between the emitted power given by
\Bekefi (valid in the cylindrical limit) and
\FullPank (including field line curvature) is not very
large if we consider only emission by a single particle. Single
particle synchrotron spectra calculated by \Bekefi and
$\FullPank\!$, as well as the approximate formulas
\PankOne and \PankTwo are shown in
Fig.~\ref{single_scans}(a) and (b) for particles with normalized
momentum $p=50$ (corresponding to a particle energy of roughly 25 MeV)
and $\vpe/\vpa=0.1$ in two different tokamaks.  For such particles,
the peak emission is for wavelengths of a few $\mu$m (the near
infrared part of the spectrum). Figure~\ref{single_scans}(a) shows
that for medium-sized tokamaks (such as DIII-D),
$\FullPank\,\!$ is closely approximated by
\PankTwo. This is not surprising, as
\PankTwo is valid in most of the wavelength range
considered (especially for shorter wavelengths), whereas
\PankOne is only valid for longer wavelengths for these
parameters. For large tokamaks (such as ITER),
\FullPank is best approximated by \Bekefi,
as the effects of field curvature become small for such large major
radii. Figure~\ref{single_scans}(b) shows that \PankTwo
is not a good approximation in this case, which is expected, since
\PankTwo is not valid in this region.


Figure~\ref{single_scans}(c) and (d) investigate the energy dependence of
the above conclusions. The quantity plotted is
$\log_{10}(\mathcal{P}_i(\lambda))$.  Figure~\ref{single_scans}(c) confirms
that \PankTwo is a good approximation to
\FullPank in DIII-D for a wide range of runaway
energies. For the highest energies, agreement is still very good for
short wavelengths, but less so for longer wavelengths. This agrees
with Fig.~\ref{lambdabounds2}(a), which indicates that
\PankTwo is no longer valid for high energies and long
wavelengths. Figure \ref{single_scans}(c) also shows that for a tokamak this size, the difference between
\Bekefi and \FullPank increases with $p$, and
using \Bekefi is not recommended if quantitative
agreement is sought. In an ITER-like device, however,
Fig.~\ref{single_scans}(d) indicates that \Bekefi
approximates \FullPank very well over the whole energy
range considered. Formally, $\mathcal{P}_{\mbox{\scriptsize full}}$ reduces to $\mathcal{P}_{\mbox{\scriptsize cyl}}$ when $R \to \infty$ and $\gamma_\parallel \simeq c/v_\bot$ (where this latter relation is equivalent to $\gamma v_\bot / c \gg 1$).

\section{Spectrum from runaway electron distributions}\label{sec:distribution}
In Refs.~\cite{jaspers,yu} the synchrotron spectrum is calculated by
multiplying the single particle spectrum by the number of runaways
with a specific pitch-angle and momentum. In this section we investigate how the synchrotron spectrum changes if we take into account the whole runaway electron distribution instead of the single particle approximation considered above. We calculate the synchrotron emission integrated over a runaway electron
distribution using
\begin{equation}
P(\lambda)=\frac{2\pi}{n_{r}}\int_{R_r}f_{RE}(p,\chi)\, \mathcal{P}_i(p,\chi,\lambda)\, p^{2}\mbox{d}p\,\mbox{d}\chi\ , \label{eq:emission}
\end{equation}
where $f_{RE}$ is the runaway distribution function, $\P_i$ is one of the
single particle emission formulas discussed in the previous section,
$\chi=p_\parallel/p$ is the cosine of the pitch-angle and $n_r$ is the
runaway electron density.
The runaway region of momentum space $R_r$
is defined by a separatrix \mbox{$p_s=(\E-1)^{-1/2}$} such that
all particles with $p>p_s$ are considered runaways \cite{smith}. Here,
$\E = E_\|/E_c$ is the parallel electric field $E_\|$ normalized to the critical field
$E_c=m_e c/(e\tau)$, with $ \tau=(4\pi r_{e}^{2}n_{e}c\ln\Lambda)^{-1}
$ the collision time for relativistic electrons, $r_e$ the classical
electron radius, $n_e$ the electron density and $\ln \Lambda$ the
Coulomb logarithm. 
 As we normalize to $n_r$, $P(\lambda)$ is
the average emission per runaway.
The alternative choice of normalizing by the runaway current $I_r$
was also considered, and it was found that all results presented
below are essentially unchanged aside from an overall scale factor,
since the speed of all runaways is nearly $c$.

In large tokamak disruptions, secondary runaway generation is
expected to dominate over primary generation,
in which case
the runaway distribution will grow approximately exponentially in time:
$\partial f_{RE}/\partial t \propto f_{RE}$.
In this case of exponential growth, the
electron distribution can be approximated
by~\cite{fulop2006}:
\begin{equation}
f_{RE}(p_{\parallel},p_{\perp})=\frac{n_{r}\hat{E}}{2\pi c_{z}p_{\parallel}\ln\Lambda}\,\exp\left(-\frac{p_{\parallel}}{c_{z}\ln\Lambda}-\frac{\hat{E}p_{\perp}^{2}}{2p_{\parallel}}\right), \label{eq:anal_ava}
\end{equation}



\noindent where $\hat{E}=(\E-1)/(1+\zeff)$, $\zeff$ is the effective
ion charge and $c_{z}=\sqrt{3(\zeff +5)/\pi}$, and the momentum space coordinates are related to $p$ and $\chi$ through $p_\|=p\chi$ and \mbox{$p_\perp = p\sqrt{1-\chi^2}$}. 
Derivation of Eq.~(\ref{eq:anal_ava}) assumes strong anisotropy ($p_\perp \ll p_\|$) and high electric field ($\bar{E}\gg 1$).
In addition to the lower boundary $p=p_s$ of the runaway region, an
upper cut-off $p=\pmax$ of the distribution will be introduced. This cut-off is
physically motivated by the finite life-time of the accelerating
electric field and the presence of loss mechanisms such as radiation
and radial transport.

\begin{figure}
\begin{center}
{\includegraphics[width=0.48\textwidth]{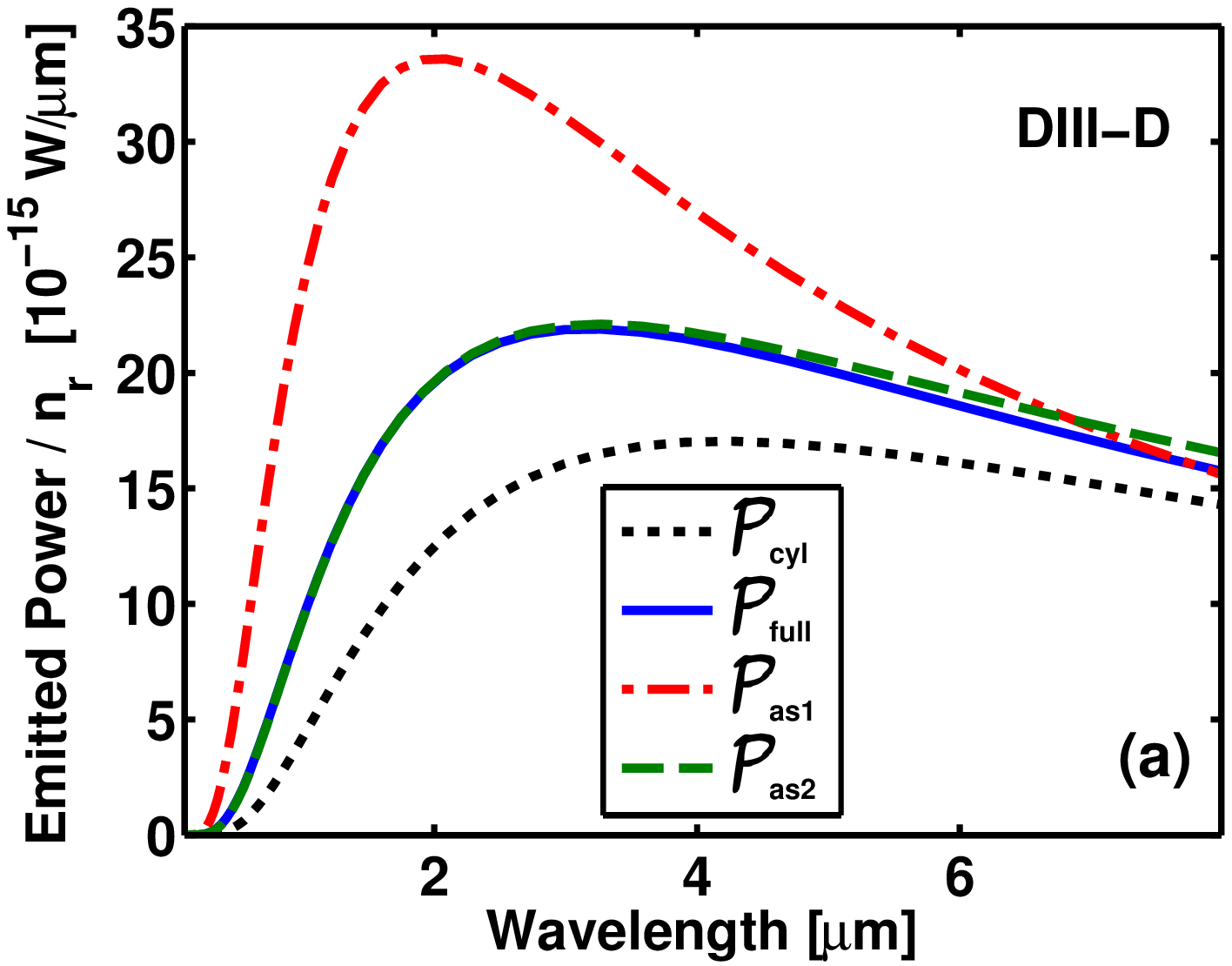}}{\includegraphics[width=0.48\textwidth]{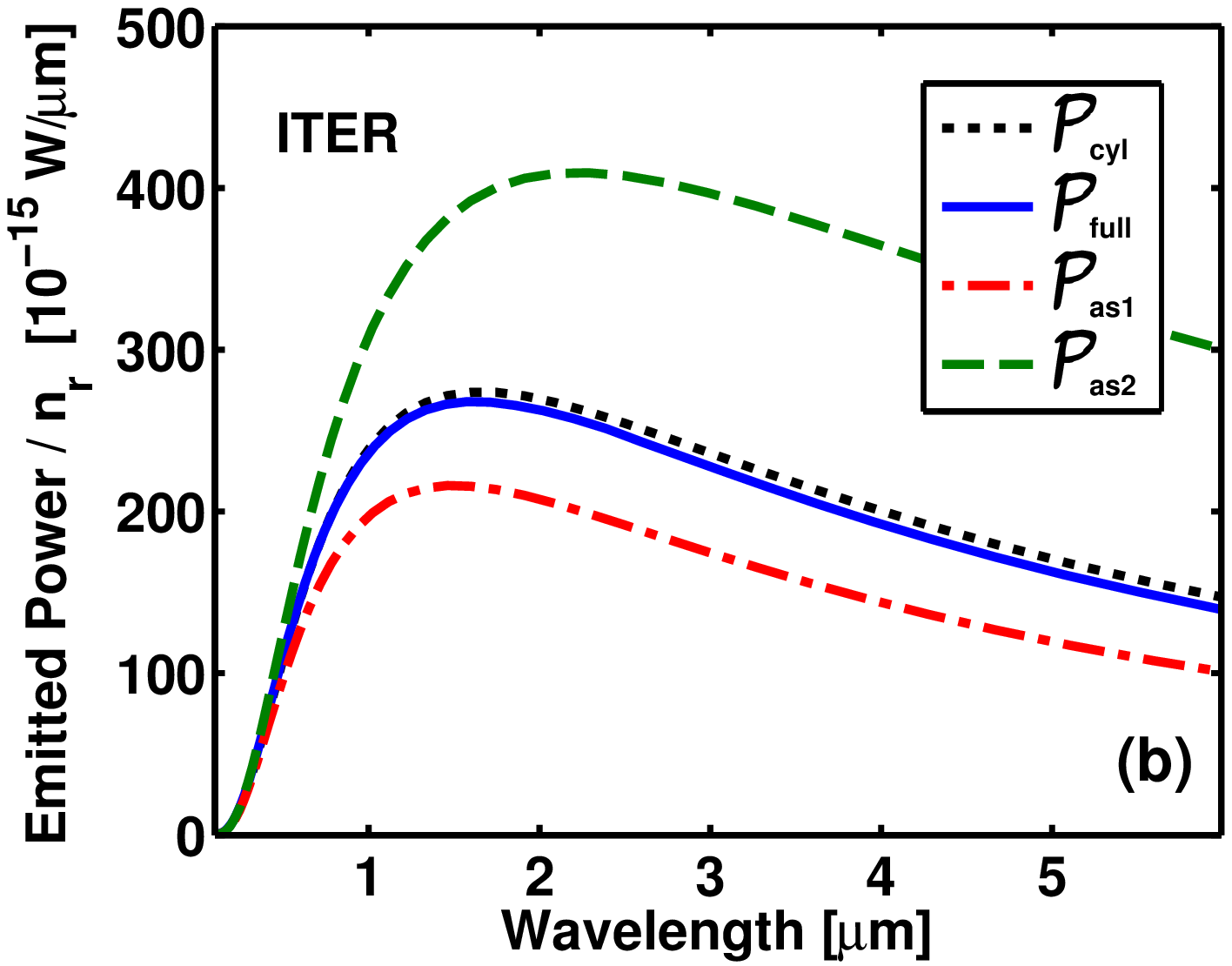}}
\caption{(Color online) Comparison of the synchrotron spectrum from a runaway distribution (Eq.~\ref{eq:anal_ava}), as calculated using \Bekefi, \FullPank, \PankOne
  or \PankTwo. 
Normalizing the emitted power by the runaway current $I_r$ instead of by $n_r$ gives negligible difference
in these figures or any figures below (the curves
are not even distinguishable), 
since most runaways move at speed $\approx c$.
  }
\label{fig:approx_formulas}
\end{center}
\end{figure}

As it was shown in the previous section, the inclusion of field
curvature effects via the use of \FullPank rather than
\Bekefi had little effect on the synchrotron emission of
a single particle in an ITER-sized device. The effect is larger in
smaller devices. When the complete runaway distribution is taken into
account, these conclusions still
hold. Figure~\ref{fig:approx_formulas} shows synchrotron spectra
calculated using Eq.~(\ref{eq:emission}) together with the
distribution in Eq.~(\ref{eq:anal_ava}) and the emission formulas
\Bekefi, \FullPank, \PankOne and
\PankTwo. The calculation was performed for both a
DIII-D-size and an ITER-size device, as the field curvature is what
separates the different formulas. The parameters used in the
calculation in Fig.~\ref{fig:approx_formulas} are maximum normalized
momentum $\pmax=100$ (corresponding to a maximum runaway energy of
roughly 50 MeV), parallel electric field $E_{\parallel}=2\,$V/m,
effective charge $\zeff=1$, background electron density
$n_{e}=3\power{20}\,$m$^{-3}$ and background plasma temperature
$T=10\,$eV. The relatively low temperature is what is expected after a
thermal quench in a disruption. In DIII-D, the post thermal-quench
temperature is estimated to be as low as $T=2$ eV
\cite{hollmann}.

Figure~\ref{fig:approx_formulas}(a) shows that in DIII-D,
\FullPank is well approximated by \PankTwo,
especially in the short wavelength slope region of the spectrum. In
ITER, \Bekefi is a good approximation, as shown in
Fig.~\ref{fig:approx_formulas}(b). This is expected since the field
curvature is much smaller here. These results are consistent with the
conclusion regarding single particles in
Fig.~\ref{single_scans}. 
For simplicity, throughout the remainder of this paper we will use \Bekefi when
calculating synchrotron spectra (except for the comparison with DIII-D
data in Sec.~\ref{sec:DIII-D}). Synchrotron spectra
calculated by \Bekefi and \FullPank are
qualitatively similar for both small and large machines, and are also often quantitatively similar
for large machines.


\begin{figure}
{\includegraphics[width=0.46\textwidth]{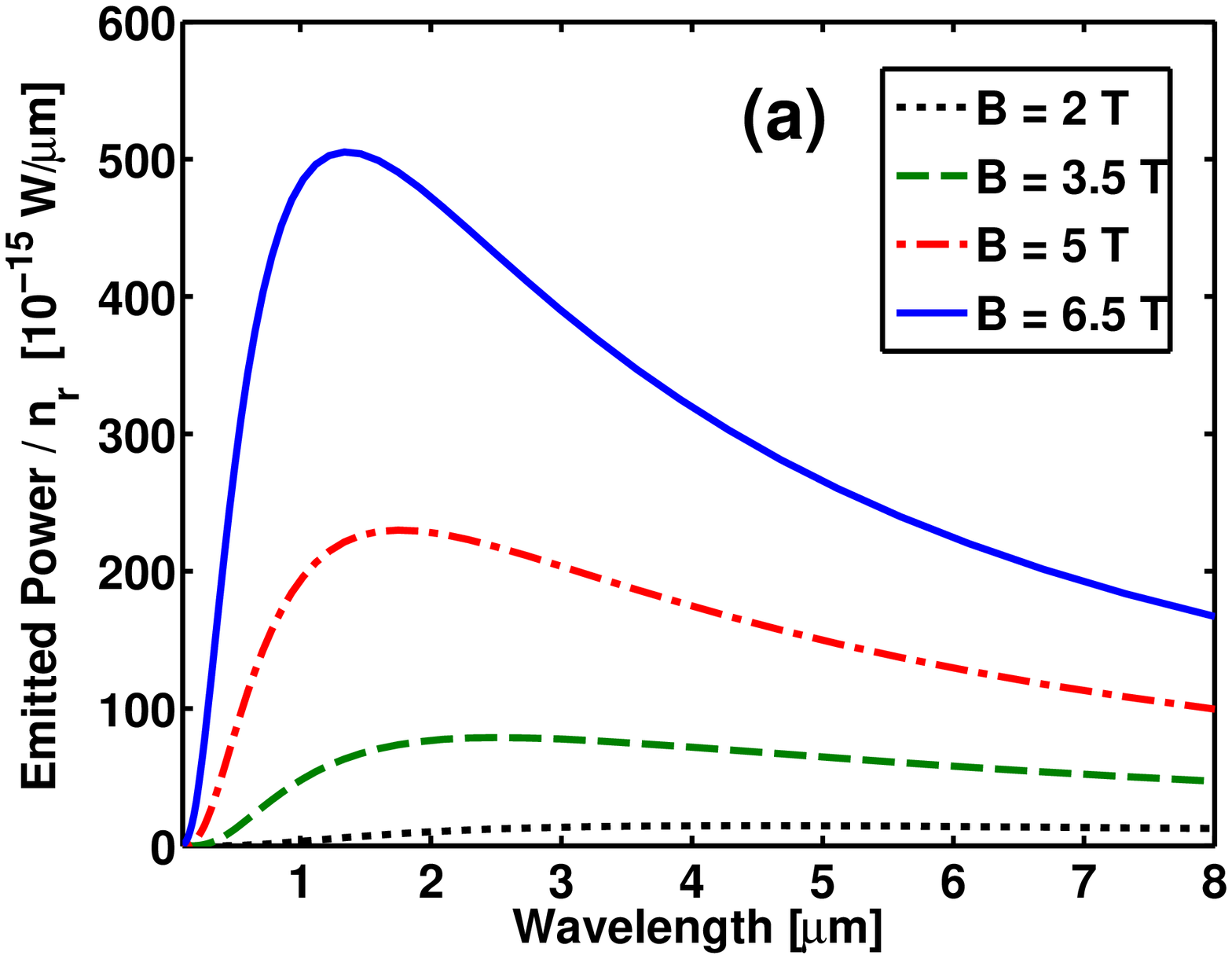}} 
{\includegraphics[trim=0cm 0cm 1.3cm 0cm, clip=true,width=0.46\textwidth]{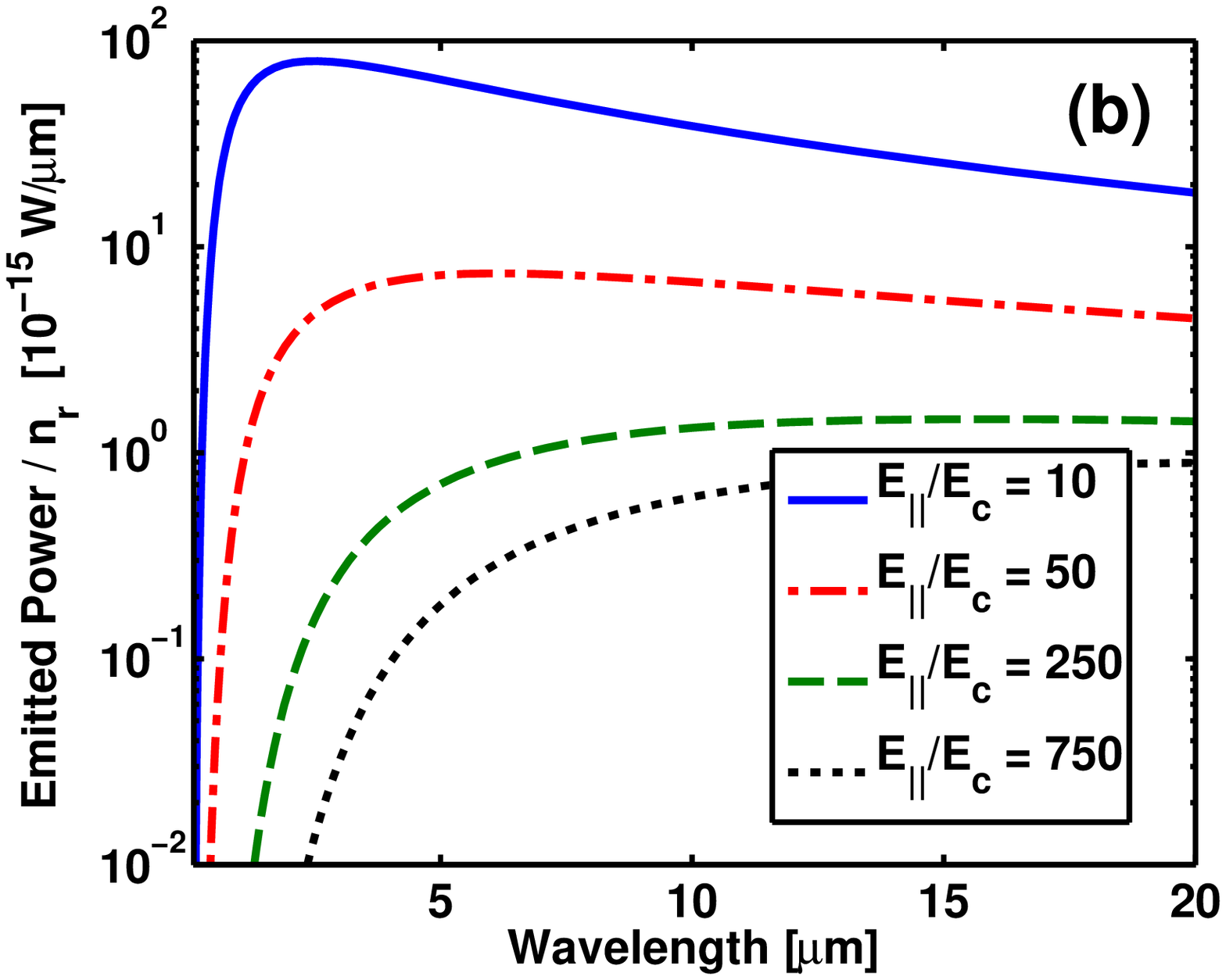}}\\
{\includegraphics[width=0.46\textwidth]{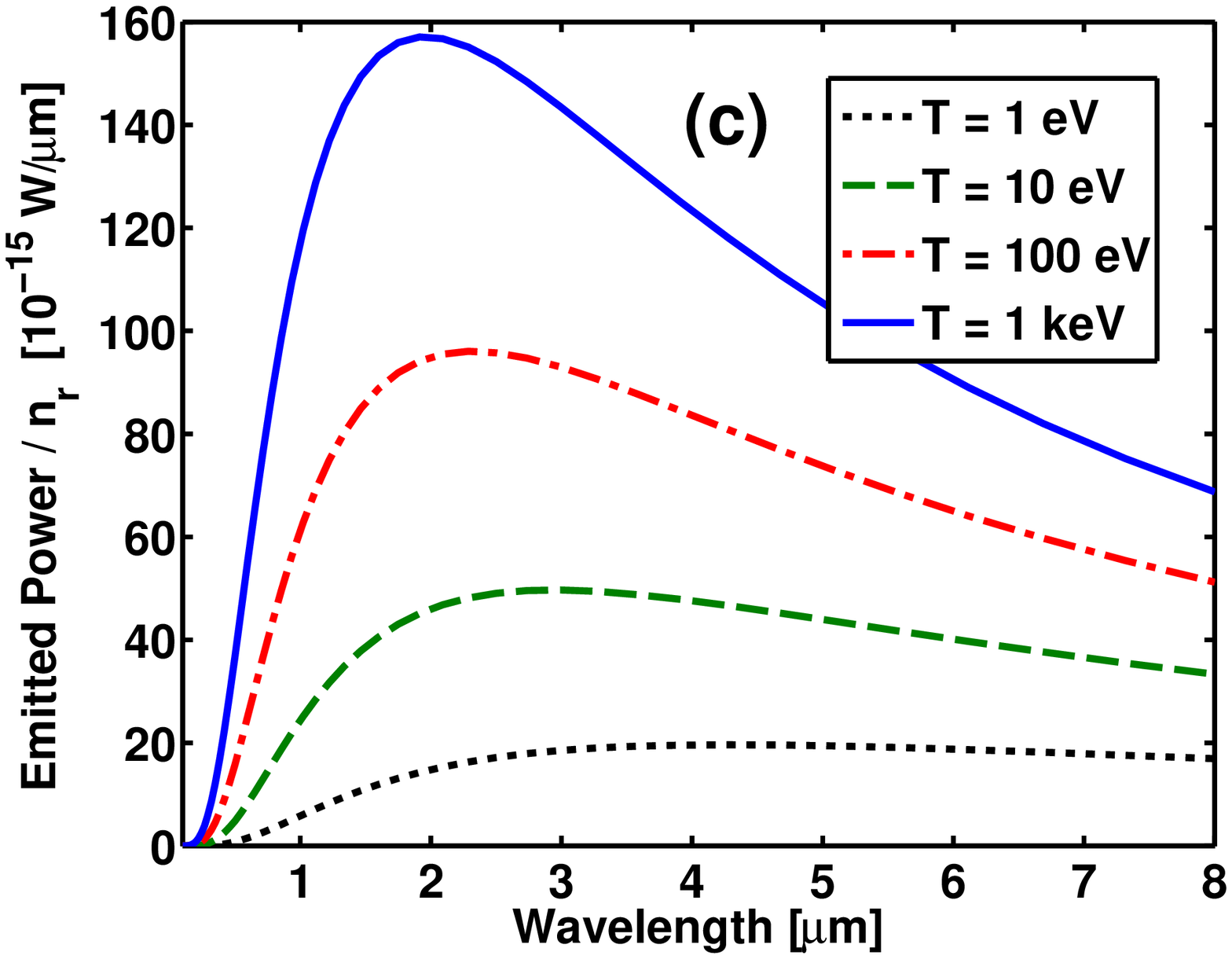}}
{\includegraphics[width=0.46\textwidth]{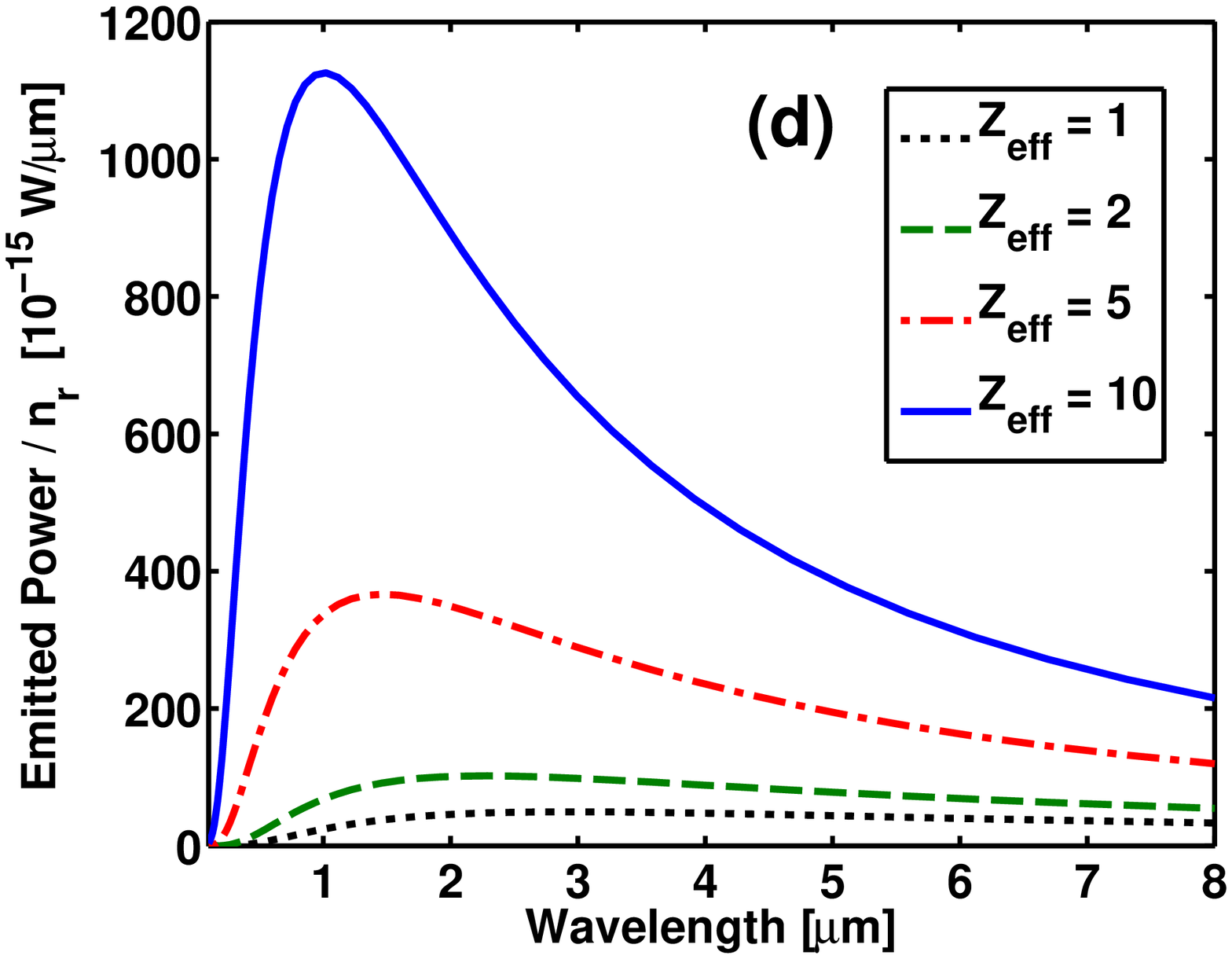}}\\
{\includegraphics[width=0.46\textwidth]{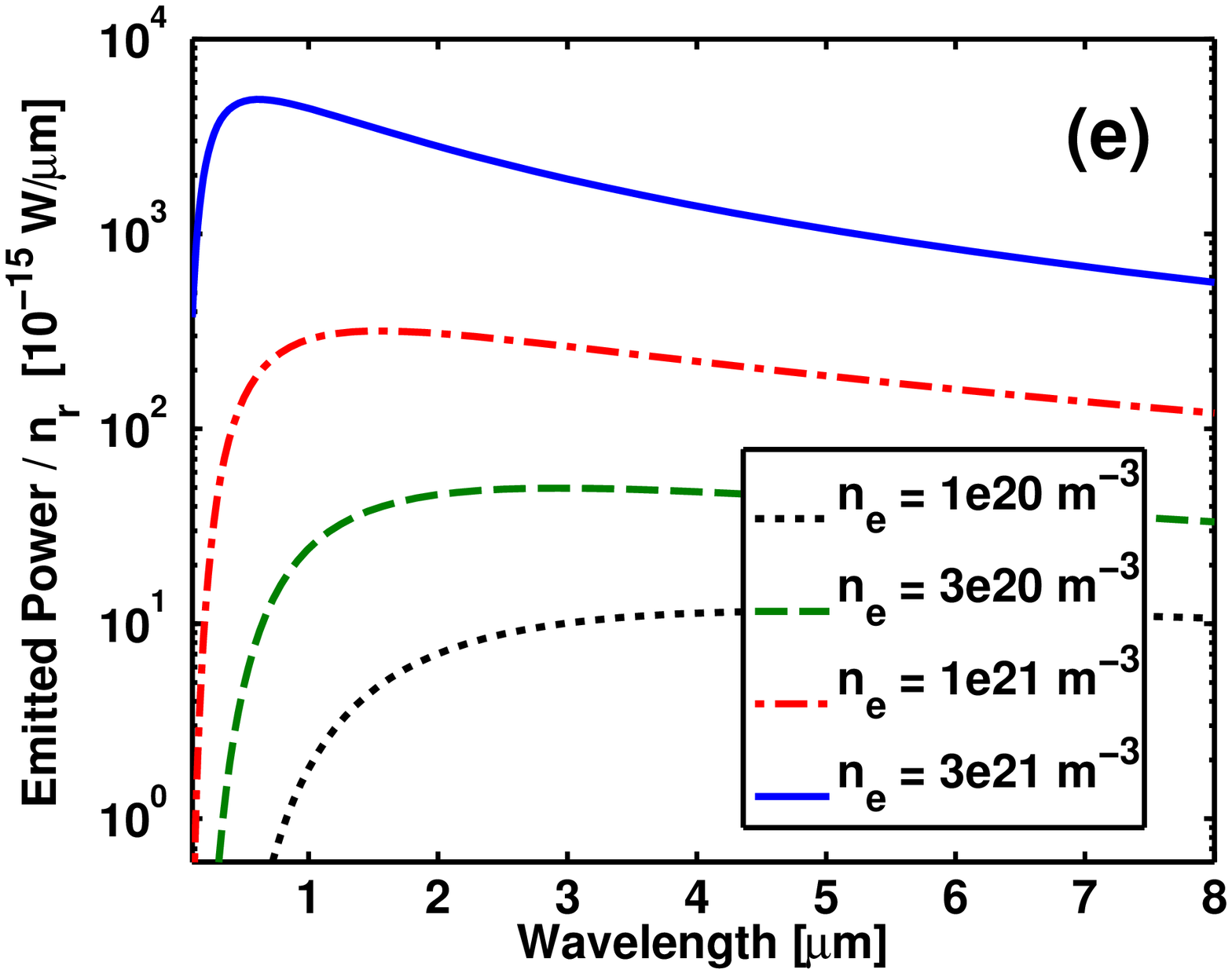}}
{\includegraphics[width=0.46\textwidth]{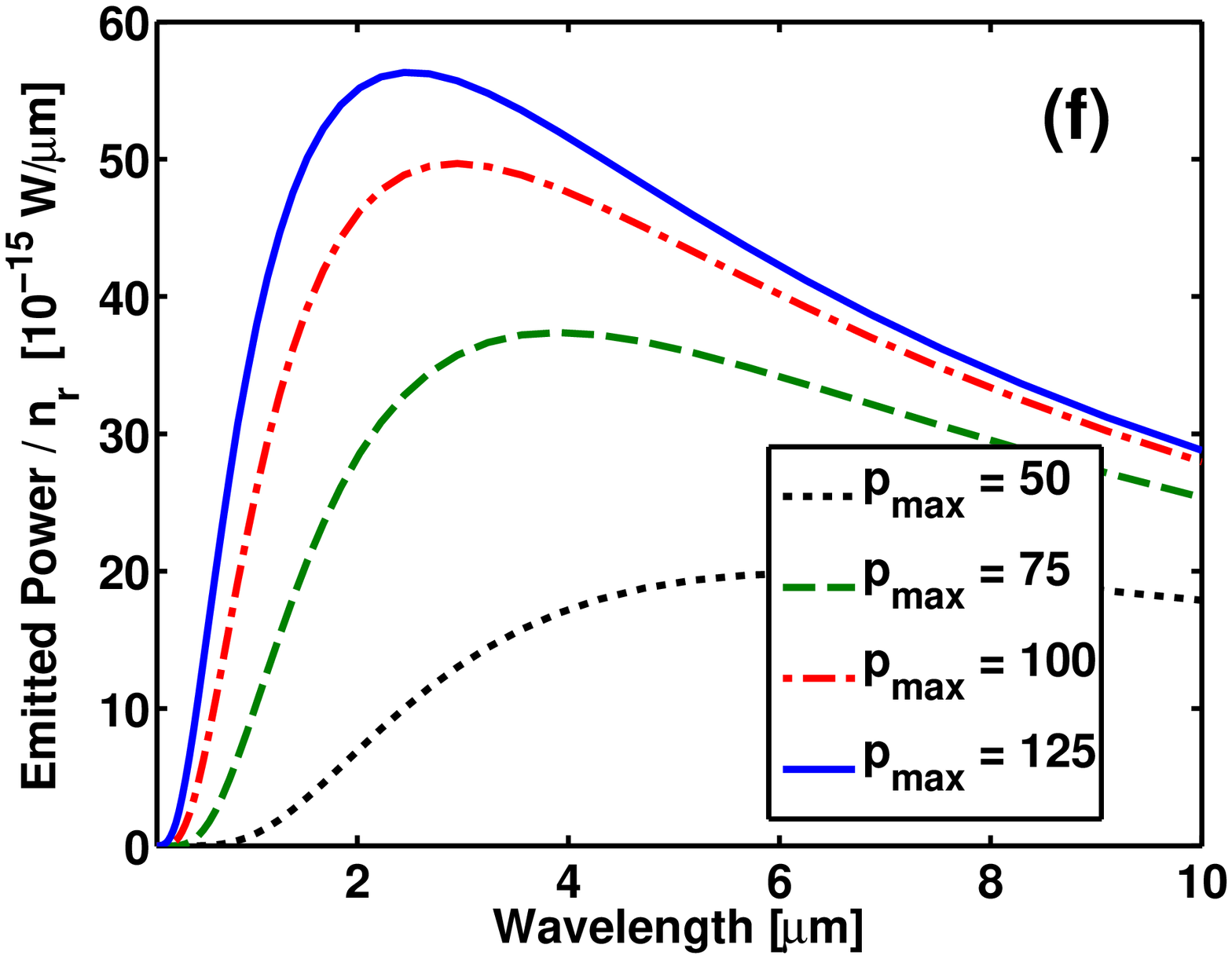}}
\caption{(Color online) Synchrotron spectra calculated using Eq.~(\ref{eq:emission}) together with \Bekefi and Eq.~(\ref{eq:anal_ava}). Note that the spectra are normalized to the runaway density. Unless otherwise noted, the parameters are $\pmax=100$, 
  $E_{\parallel}=2\,$V/m, $\zeff=1$, \mbox{$n_{e}=3\power{20}\,$m$^{-3}$},
  $T=10\,$eV, and $B = 3\,$T. For this scenario, $E_c = 0.15\,$V/m. } \label{scans}
\end{figure}

The single particle synchrotron emission formulas are independent of
the plasma temperature, effective charge, density and the strength of
the electric field.  These quantities do however affect the shape of
the runaway distribution, which in turn affects the synchrotron
emission. Figure~\ref{scans} shows scans in these parameters, the
magnetic field and maximum momentum $p_{max}$ of the distribution. The
baseline scenario corresponds to the parameters used in
Fig.~\ref{fig:approx_formulas} together with $B=3\,$T. Since \Bekefi is used, there is no dependence on $R$.

Figure~\ref{scans} shows that the average synchrotron emission
increases with $B$, $T$, $\zeff$, $n_e$ and $\pmax$, but decreases
with increasing electric field strength. 
The dependence on $n_e$ and $\E$ is particularly strong, and we note
that the average emission can vary over several orders of
magnitude. This variation is completely missing from the single
particle approximation used in Section \ref{sec:spectrum}. If, as a
disruption mitigation technique, a large amount of material is injected
into the plasma (for instance in the form of a massive gas injection),
the increase in density would lead to increased synchrotron emission
from the runaways (if the mitigation is unsuccessful). This could give
the impression of an increase in the number of runaways even though
this is not necessarily the case.  The figure also shows that the
wavelength of peak emission shifts appreciably with varying parameter
values. Generally, an increased average emission is accompanied by a
shift of the peak emission towards shorter wavelengths. The total
synchrotron emission of a single particle scales roughly as
$(\gamma\vpe/\vpa)^2$
\cite{jaspers}. 
Thus, the most strongly emitting particles are highly energetic with
large pitch-angle. These particles emit at shorter wavelengths, so the
shift of the wavelength of peak emission with increased total emission
is
expected. 



\begin{figure}
{\includegraphics[width=\textwidth]{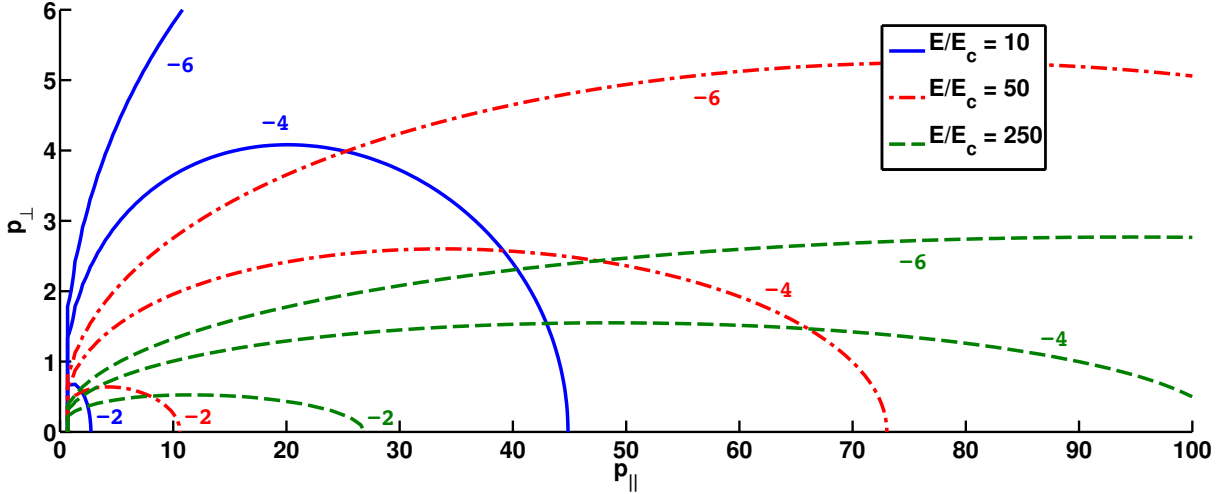}}

\caption{(Color online) Shape of the analytical avalanche distribution
  (Eq. \ref{eq:anal_ava}) for three of the parameter sets in
  Fig.~\ref{scans}(b). The plot shows contours of the quantity
  $\log_{10}|f_{RE} / n_r|$. }
\label{fig:contours}
\end{figure}

In light of the particle energy dependence of the emitted synchrotron
power, the decrease in emission with increasing electric field
strength may seem a little surprising, as a stronger accelerating
field leads to more highly energetic particles. The explanation can be
found in the shape of the runaway beam. Figure~\ref{fig:contours}
shows the runaway distribution, Eq.~(\ref{eq:anal_ava}), in
($\ppa,\pp$)-space for three of the parameter sets in the electric
field scan in Fig.~\ref{scans}(b). The figure shows that the
distribution, in addition to being extended in $\ppa$, becomes more
narrow in $\pp$ as the electric field strength increases. This leads
to lower average-per-particle emission by virtue of the pitch-angle dependence of
\Bekefi, despite the presence of a greater number of
highly energetic particles.

%

\begin{figure}
\begin{center}
{\includegraphics[width=0.49\textwidth]{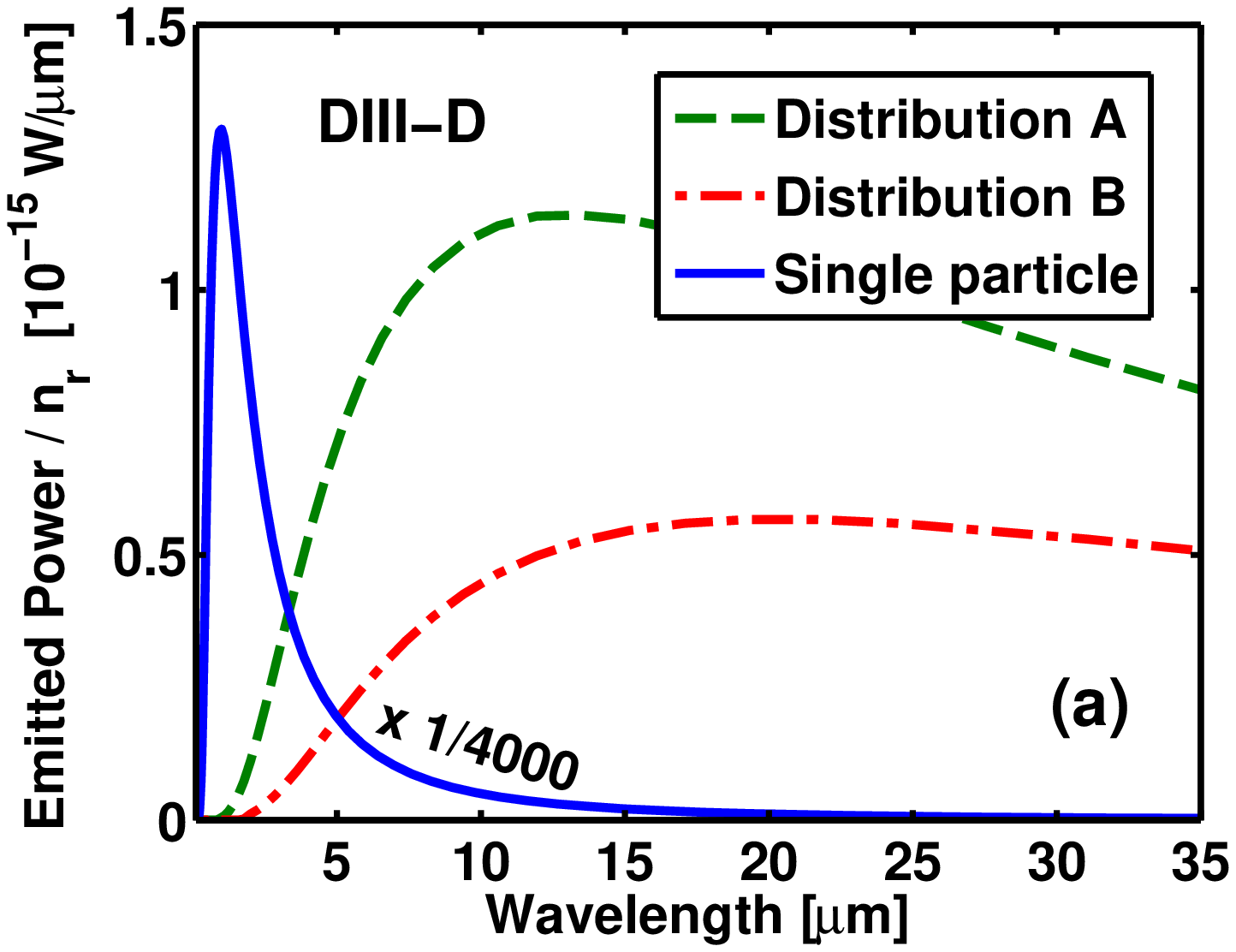}}
{\includegraphics[width=0.49\textwidth]{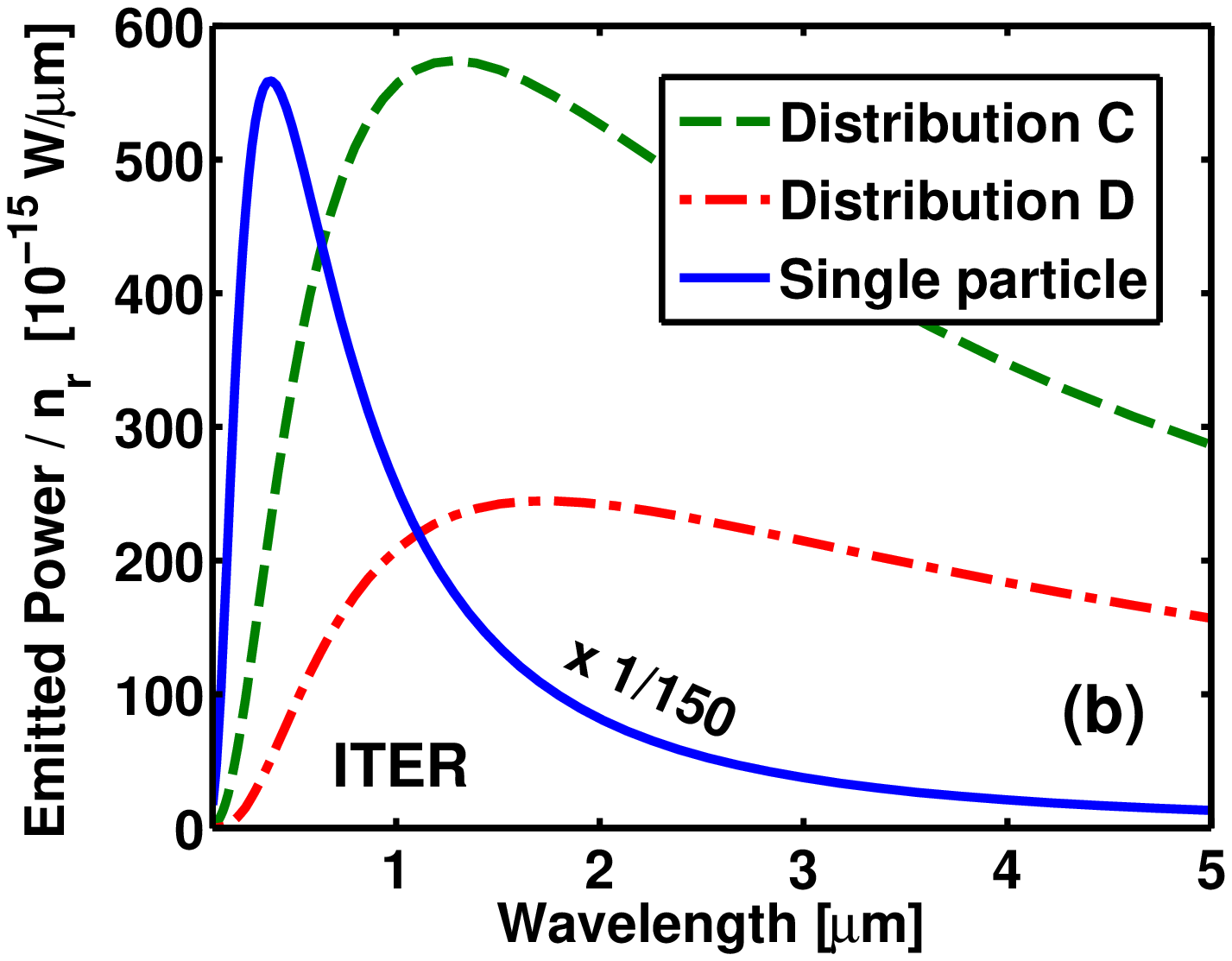}}
\caption{(Color online) Synchrotron spectra (average emission per particle) calculated using the runaway distribution in Eq.~\eqref{eq:anal_ava} and \Bekefi for DIII-D-like and ITER-like cases. The synchrotron spectrum from a single particle
  with $p=100$ and $v_\perp/v_\|=0.15$ is also shown. Note that the single particle spectra have been multiplied by a small factor to fit on the same scale. The parameters used for the distributions were $\pmax=100$ and 
  {\bf A}: $E_{\parallel}=2\,$V/m, $\zeff=1$, \mbox{$n_{e}=5\power{19}\,$m$^{-3}$},
  $T=2\,$eV, 
  {\bf B}: $E_{\parallel}=10\,$V/m, $\zeff=1.5$, \mbox{$n_{e}=1\power{20}\,$m$^{-3}$},
  $T=2\,$eV, 
  {\bf C}: $E_{\parallel}=2\,$V/m, $\zeff=1$, \mbox{$n_{e}=5\power{20}\,$m$^{-3}$},
  $T=10\,$eV, 
  {\bf D}: $E_{\parallel}=10\,$V/m, $\zeff=2$, \mbox{$n_{e}=1\power{21}\,$m$^{-3}$},
  $T=10\,$eV.}
\label{fig:comparison}
\end{center}
\end{figure}

Figure \ref{fig:comparison} shows a comparison of the average synchrotron
spectrum calculated for the runaway distribution Eq.~(\ref{eq:anal_ava}) and for a single
particle. 
The figure clearly shows that using the single-particle emission
overestimates the synchrotron emission per particle by several orders
of magnitude.  (Note that the values for the emitted power per
particle were divided by a large number to fit in the same scale.) The
overestimation is caused by the fact that the single-particle
approximation assumes that all particles emit as much synchrotron
radiation as the most strongly emitting particle in the actual
distribution, as discussed in Section \ref{sec:introduction}. Furthermore,
the wavelength of peak emission is shifted towards shorter wavelengths
when using this approximation. Using the single-particle approximation can thus give misleading results regarding both the spectrum shape and the total emission strength.

\section{Synchrotron radiation as a runaway electron diagnostic}
\label{sec:discussion} 
The interest in the synchrotron emission of runaways is primarily
motivated by its potential as a runaway diagnostic. 
In principle, the distribution can be determined by acquiring
an experimental synchrotron spectrum and comparing it to calculations
using Eq.~\eqref{eq:emission} for a range of $\pmax$, provided all
other relevant parameters are known. There are however several
problems with this approach. First, the complete synchrotron
spectrum is not known. Detectors are only sensitive in a limited
wavelength range, which is likely to also contain contaminating
radiation from other sources in the plasma. Second, the relevant
plasma parameters are not always well known, especially during
disruptions. This can lead to significant uncertainty in the computed
synchrotron spectrum, as the parameter scans in Fig.~\ref{scans}
indicated. Using a single particle approximation for the runaway
distribution seemingly avoids the second issue, but as we have seen,
it also ignores factors that can influence the emission by orders of
magnitude.

\subsection{Spectrum slope  and maximum runaway energy}
\begin{figure}
\includegraphics[width=0.49\textwidth]{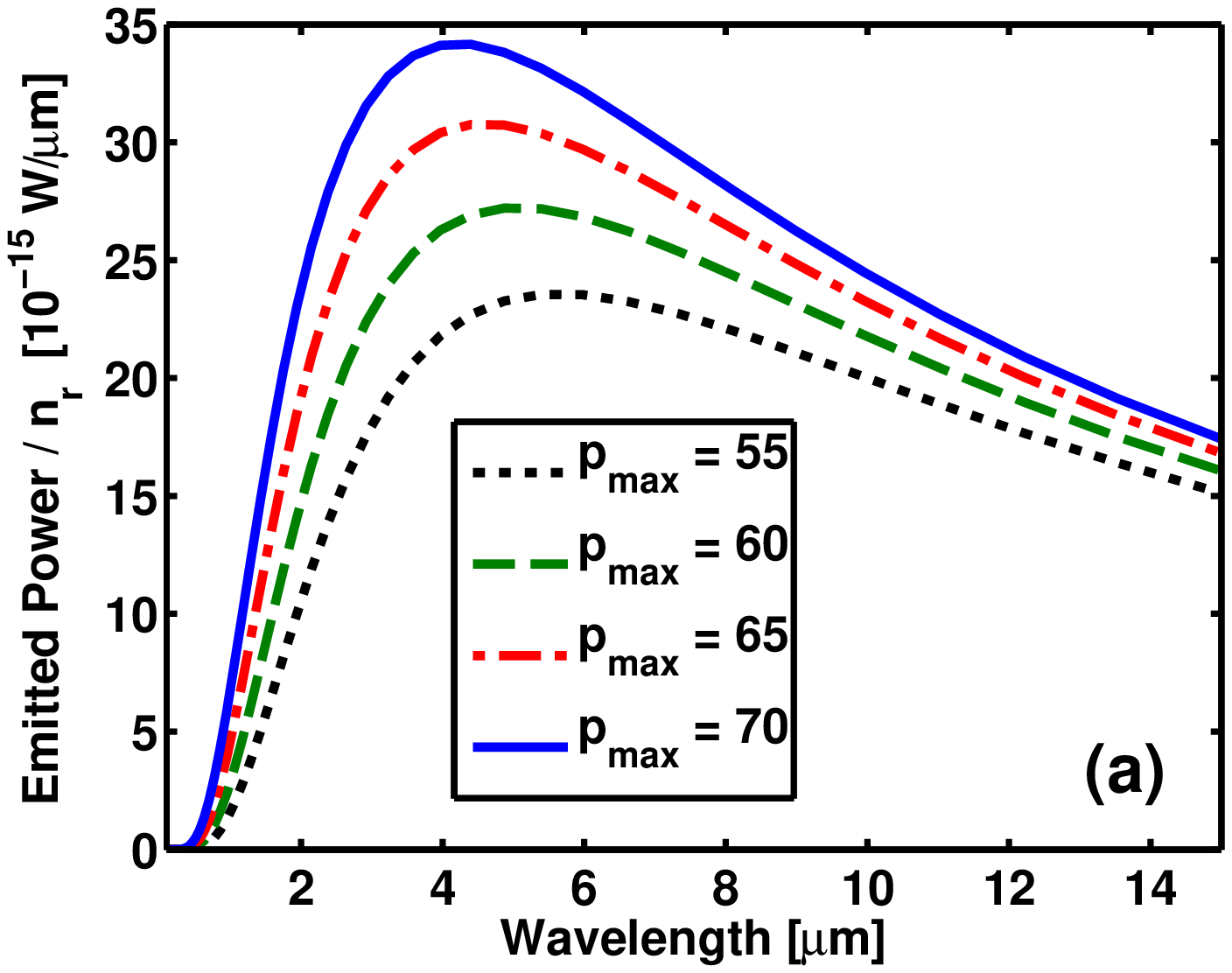}
\includegraphics[width=0.49\textwidth]{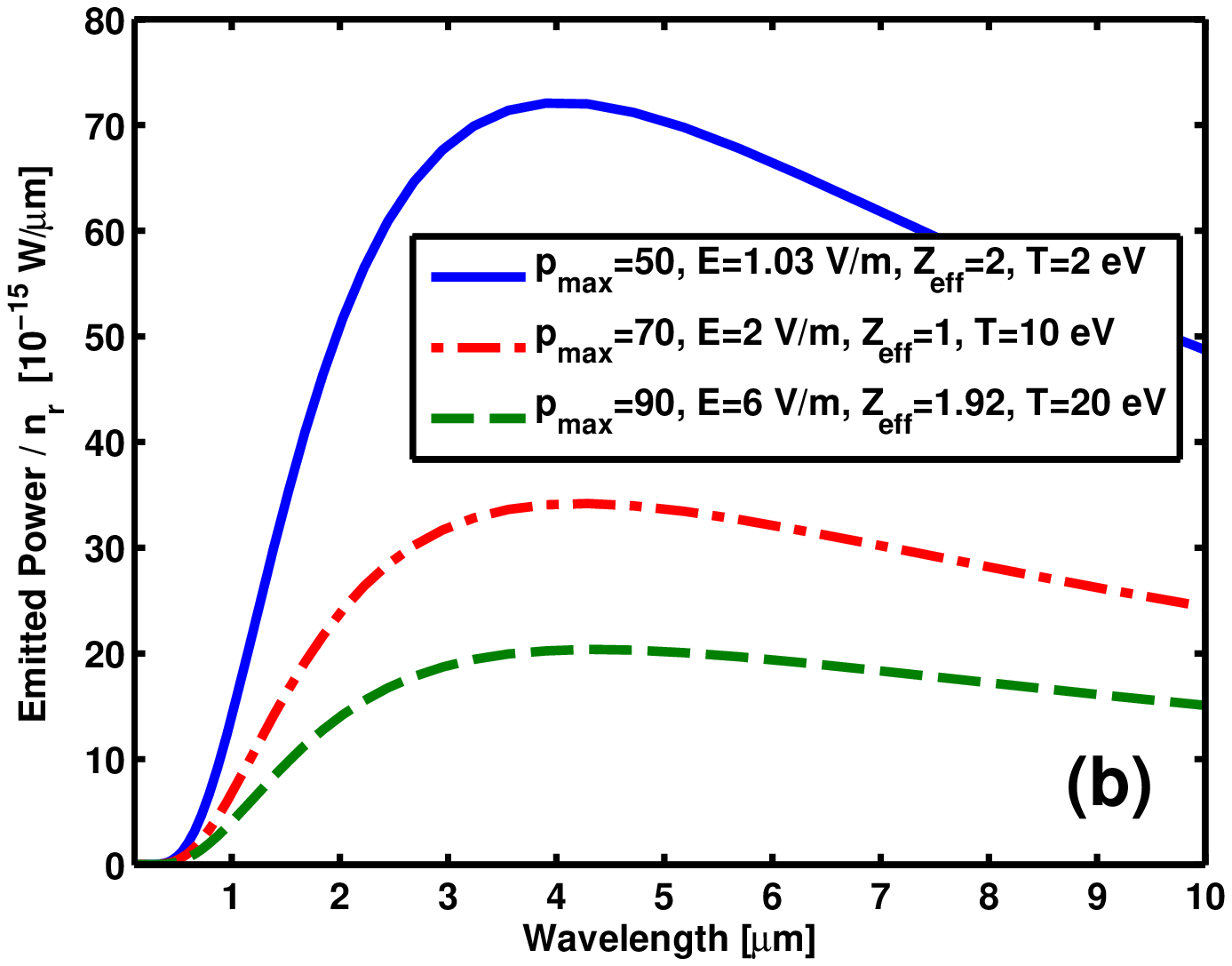}
\caption{(Color online) Spectra calculated using the analytical avalanche
  distribution Eq.~(\ref{eq:anal_ava}) and \Bekefi. In (a), the parameters used
  are the same as the baseline scenario in Fig.~\ref{scans}, but with
  different maximum particle momenta. All the curves in panel (b) have
  the same slope $S$, as calculated with $\lambda_1=1.5\,\mu$m,
  $\lambda_2=2.8\,\mu$m. The plasma parameters that differ between the
  curves are indicated in the figure. The remaining parameter values
  are $n_{e}=3\power{20}\,$m$^{-3}$, and $B=3\,$T.  }
\label{fig:slope_figures}

\end{figure}

Simple measurements of the synchrotron power for different wavelengths
on the steep slope of the spectrum have been used to estimate the
runaway energy \cite{jaspers}, using the single particle emission
formulas and assuming mono-energetic runaways with well-defined
pitch-angle. In this case there is a monotonic relationship between
the slope and the particle energy (as the wavelength of peak emission
decreases monotonically with increasing $p$). The slope can be
obtained through a relative measurement of the synchrotron power at
two wavelengths, $ S=P(\lambda_{1})/P(\lambda_{2})$.  
However, as the runaway distribution is
sensitive to the plasma parameters, when taking it into account there is in general no such simple
relationship between the slope of the spectrum and the maximum runaway
energy in the distribution. If all other parameters are fixed the
relation still holds, as is shown in
Fig.~\ref{fig:slope_figures}(a). This follows naturally from the
relation for single particles, as when $\pmax$ is increased, more
particles that emit at short wavelengths are included, and the average
emission correspondingly shifts towards shorter wavelengths, affecting
the slope.  But if the plasma parameters are uncertain, the slope can
be misleading. Figure~\ref{fig:slope_figures}(b) shows multiple
spectra with the same slope $S$ for $\lambda_1=1.5\,\mu$m and
$\lambda_2=2.8\,\mu$m. Using only a measurement of $S$ in the above
range, they cannot be distinguished, despite the appreciable
difference in average emission. This type of two-point slope
measurement can be performed using physical wavelength filters placed
in front of the detector \cite{jaspers}, in which case measurements
are constrained to specific $\lambda_1$ and $\lambda_2$ that cannot be
easily changed. The $\pmax$ of the different spectra in
Fig.~\ref{fig:slope_figures}(b) range from 50 to 90, with only modest
variation of the plasma parameters $E$, $\zeff$ and $T$ (all of which
are hard to estimate during disruptions). Thus, if the plasma
properties are uncertain, there is no clear correlation between $S$
and $\pmax$ of the distribution. Another weakness of using the slope
is the difficulty in asserting that both measurement points are
actually located on the approximately linear part of the spectrum. As
the plasma parameters change, the peak of the spectrum may shift (as
discussed in connection with Fig.~\ref{scans}). Choosing $\lambda_1$
and $\lambda_2$ that are suitable for a wide range of different
conditions (as when using physical filters) is not easy. Instead of
using the slope directly, one should calculate the emission for an
assumed beam-like distribution function (e.g. similar to
Eq.~(\ref{eq:anal_ava})), and iteratively find the $p_{max}$, which
fits the synchrotron spectrum best.

\subsection{Synchrotron emission in DIII-D}
\label{sec:DIII-D}

\begin{figure}
\includegraphics[width=0.47\textwidth]{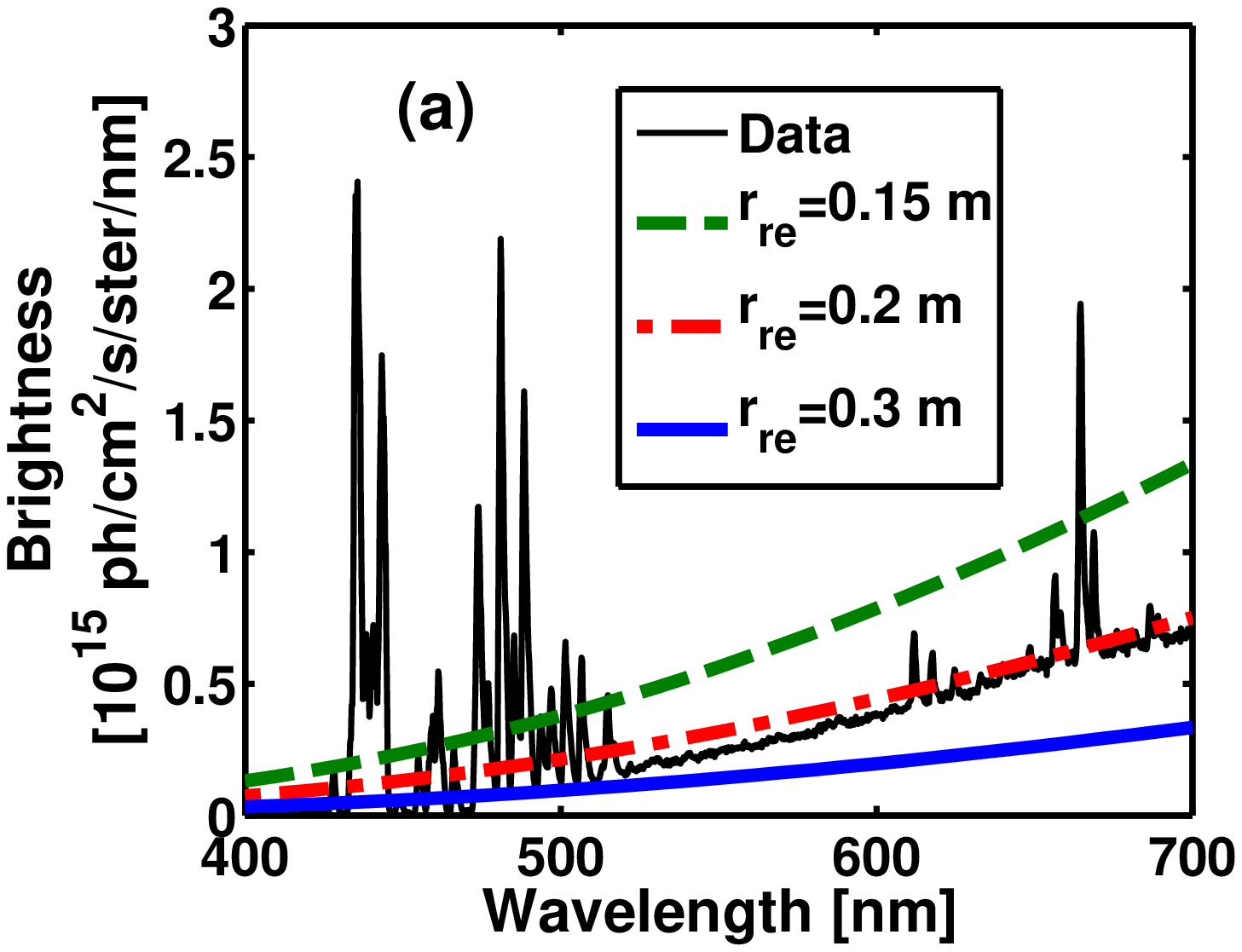}
\includegraphics[width=0.47\textwidth]{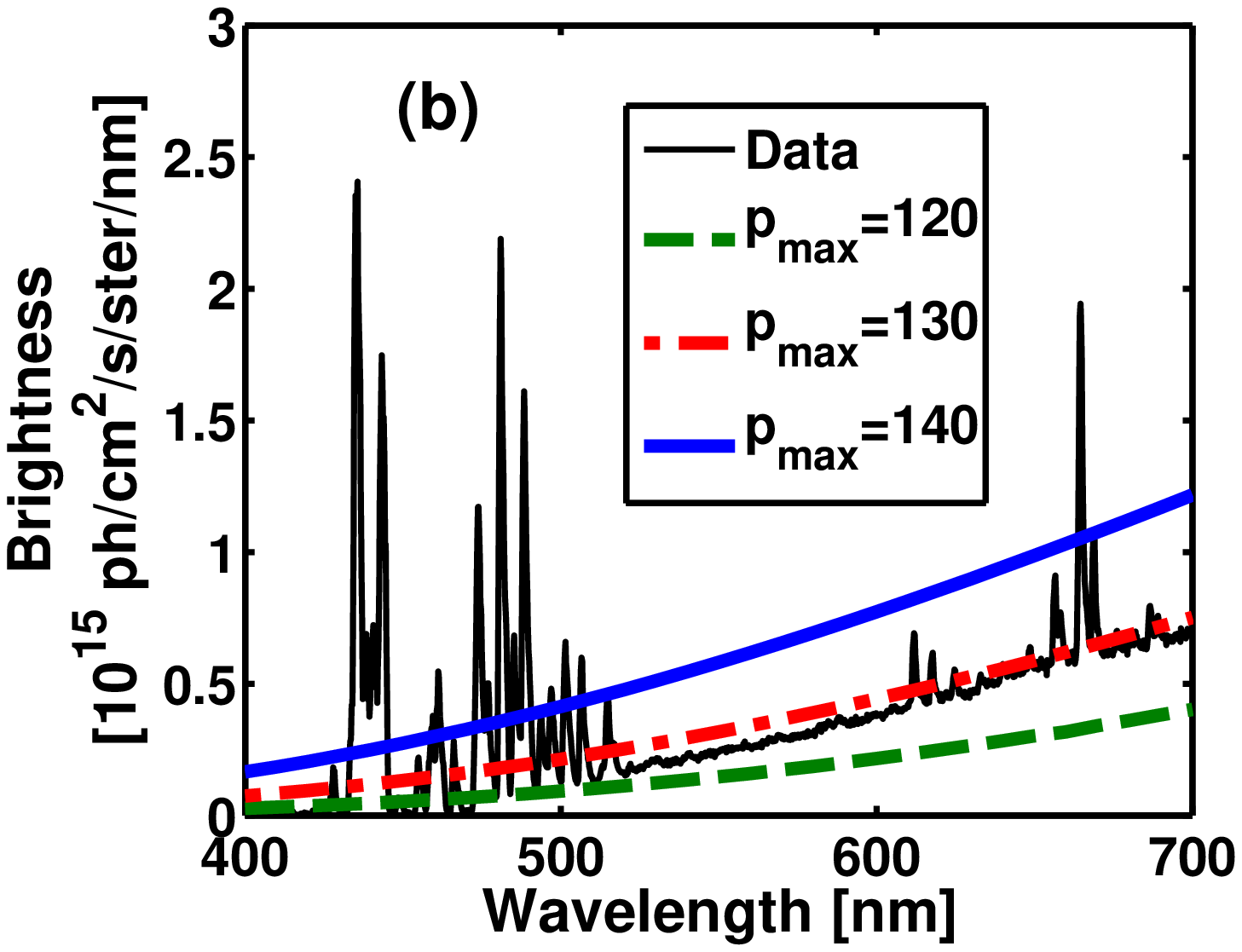}
\includegraphics[width=0.47\textwidth]{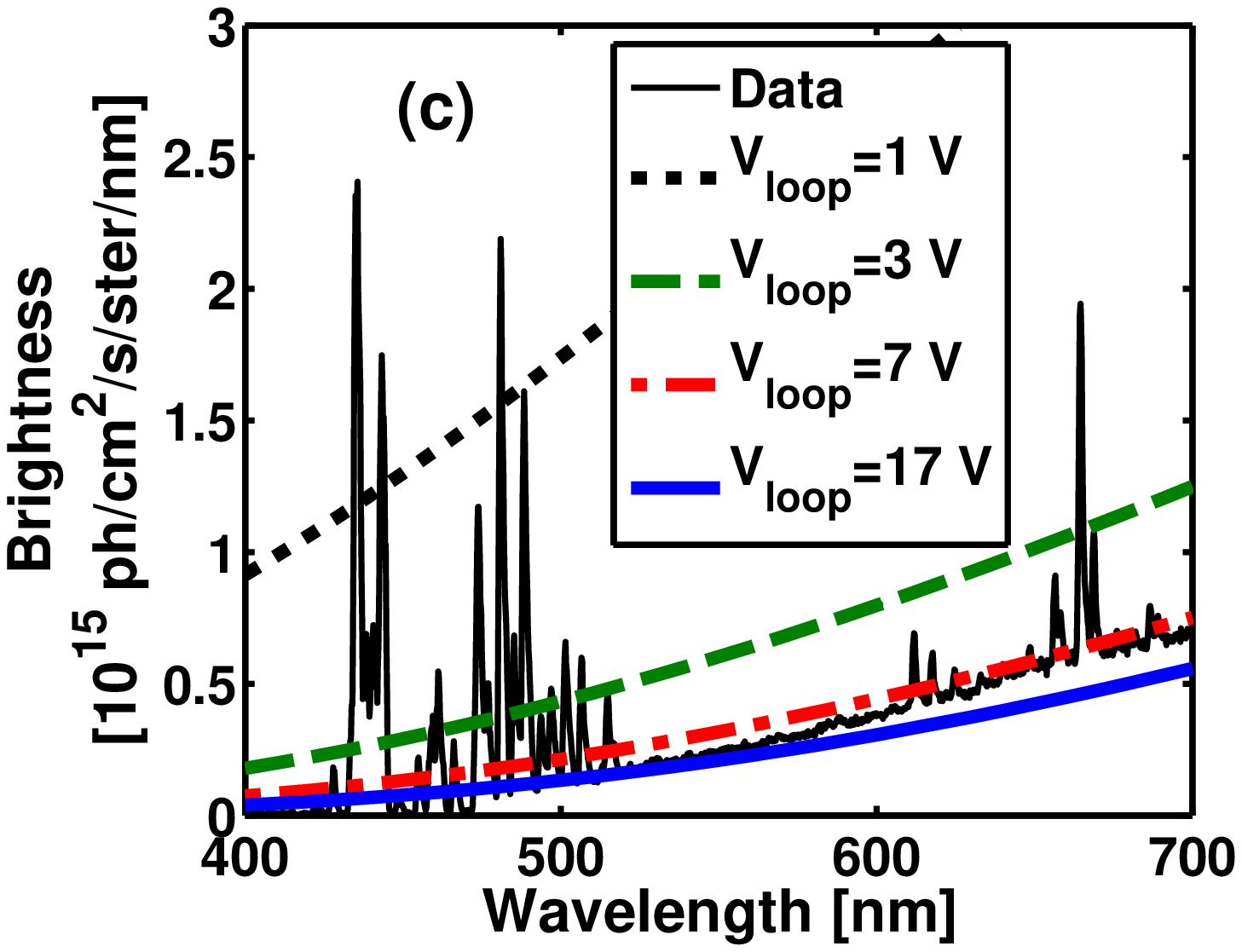}
\includegraphics[width=0.47\textwidth]{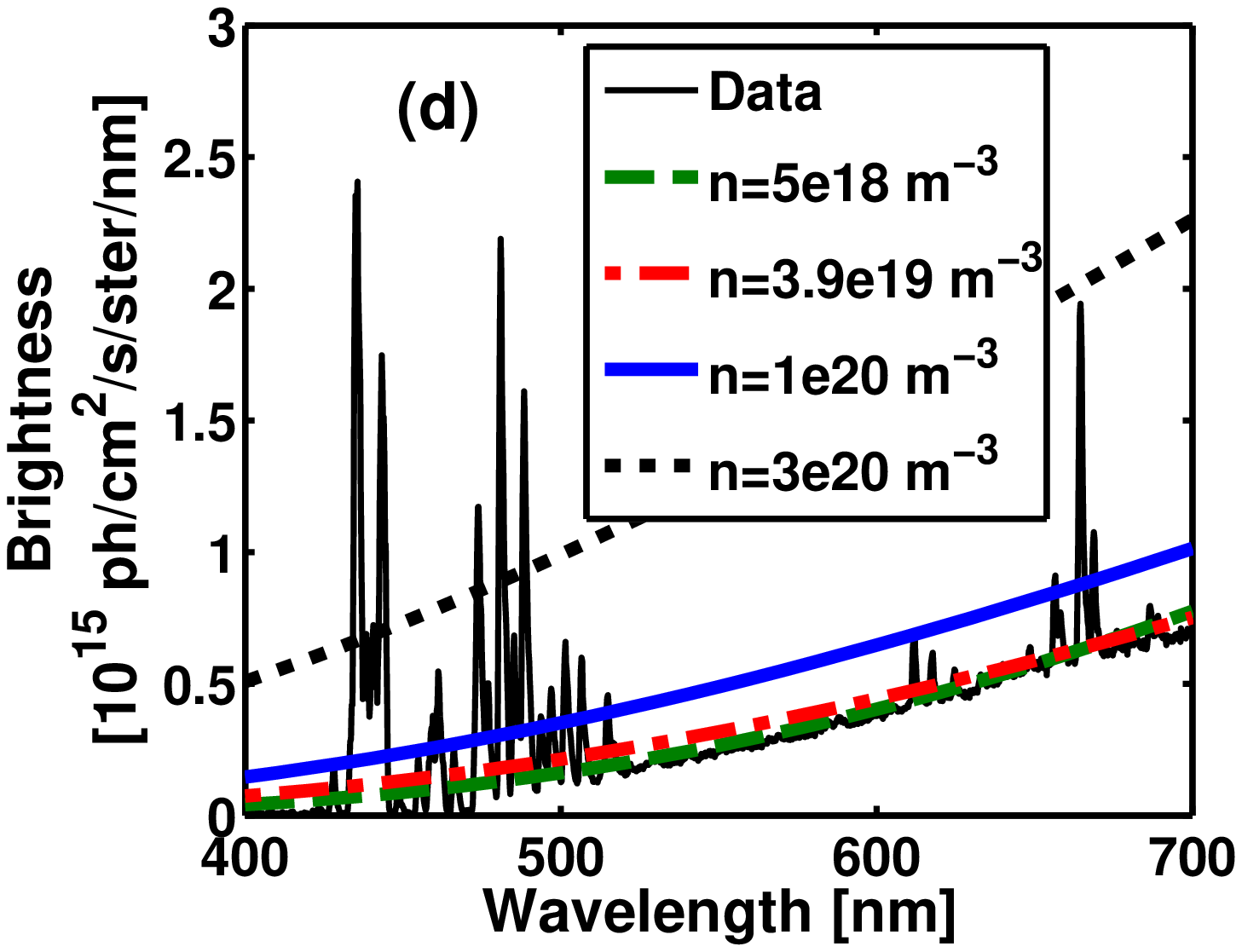}
\caption{(Color online) Measured visible spectrum in DIII-D during the
  runaway plateau at t=2290 ms in shot 146704. The data is a
  superposition of synchrotron radiation from runaways and line
  radiation from the background plasma. Theoretical synchrotron
  spectra are also shown for various (a) runaway beam radii (b)
maximum normalized momenta  $p_\text{max}$, (c) loop-voltages $V_{\text{loop}}$ and (d) densities $n$. Unless otherwise
  noted the parameters are $p_\text{max}=130$, $r_{re}=0.2 \;\rm m$,
  $n=3.9\cdot 10^{19}\;\rm m^{-3}$ and $V_\text{loop}=7\,\rm V$, which
  is indicated by the red (dash-dotted) lines. \label{fig:diiid}}
\end{figure}

It is interesting to investigate how a synchrotron spectrum calculated
for an avalanching distribution compares with an experimentally
measured synchrotron spectrum from DIII-D. In the specific
experimental scenario we consider (shot number 146704 and time t=2290
ms \cite{hollmann2013}), the loop voltage is $7 \;\rm V$, the density $3.9\cdot
10^{19}\;\rm m^{-3}$ and the plasma current $I_p=0.15\;\rm MA$,
measured near the end of a runaway plateau phase. 
The runaway density can be estimated from the current
using $n_r=I_p/(e c A_{re})$, where $A_{re}$ is the area of the
runaway beam. The runaway beam radius in this case was around 20
cm. The temperature is assumed to be $1.5\;\rm eV$ and
$Z_{\rm eff}=1$.   For synchrotron emission by mono-energetic runaway electrons the conversion to the measured brightness can be done using Eq.~(2) in Ref.~\cite{yu}:
\begin{equation}
\mathcal{B}(\lambda,\theta,\gamma)=\mathcal{P}(\lambda,\theta,\gamma)\,\frac{2R}{\pi\theta}n_{r}\,,\label{eq:brightness}
\end{equation}
where $R$ is the major radius (of the runaway beam) and $\theta=\vpvp$ is the tangent of the particle pitch-angle. Taking into
account the runaway distribution, we calculate the brightness as
\begin{equation}
  B(\lambda)=4R\int_{\chi_{min}}^{\chi_{max}}\int_{p_{min}}^{p_{max}}\frac{1}{\theta(\chi)}\,\mathcal{P}\Bigl(\lambda,\theta(\chi),\gamma(p)\Bigr)\, f(p,\chi)\, p^{2}dpd\chi\,,\label{eq:brightness_distr}
\end{equation}
where $\theta(\chi)=\tan(\arccos(\chi))=\sqrt{1-\chi^{2}}/\chi$ and
$\gamma(p)=\sqrt{p^{2}+1}$, $p_{min}=(\bar{E}-1)^{-1/2}$ and the
integration limits for the pitch-angle are $\chi_{min}=0$,
$\chi_{max}=1$. Since we consider the visible part of the spectrum, all
$p_{min}$ below $p=50$ produce identical results, as only the highest
energy particles emit in this range.  Equation (\ref{eq:brightness}) is
strictly valid for $1/\gamma\ll\theta$ \cite{yu}.  As we are
interested in the complete distribution with both small $\gamma$ and
small $\theta$, we use instead the effective viewing aperture
$\theta_{\mbox{eff}}\approx\sqrt{\theta^{2}+\gamma^{-2}+(r_{lens}/r_0)^{2}}$. Here, $r_{lens}=2$ cm is the lens aperture of the detector and $r_0\simeq2$ m is the
distance between the detector and the runaway beam.  Introducing
$\theta_{\mbox{eff}}$ into Eq.~(\ref{eq:brightness_distr}), we find
\begin{equation}
  B(\lambda)=4R\int_{R_r}\frac{1}{\theta_{\mbox{eff}}(p,\chi)}\,\mathcal{P}\left(\lambda,\chi,p\right)\, f(p,\chi)\, p^{2}dpd\chi\ .\label{eq:brightness_distr_theta_eff}
\end{equation}

Figure \ref{fig:diiid} shows a comparison of spectra calculated using Eq.~(\ref{eq:brightness_distr_theta_eff}) together with
\PankTwo and Eq.(\ref{eq:anal_ava}), and the experimentally
measured spectra for different runaway beam radii (the beam is assumed
to have circular cross-section), $p_{max}$, loop voltages and
densities. The good agreement for $r_{re}=20 \;\rm cm$ and
$p_{max}=130$ leads us to estimate the {\it maximum} runaway electron energy
to be around 65 MeV. This is much larger than  the mean energy of
several MeV estimated from other diagnostics \cite{hollmann2013}.

\begin{figure}
\includegraphics[width=0.6\textwidth]{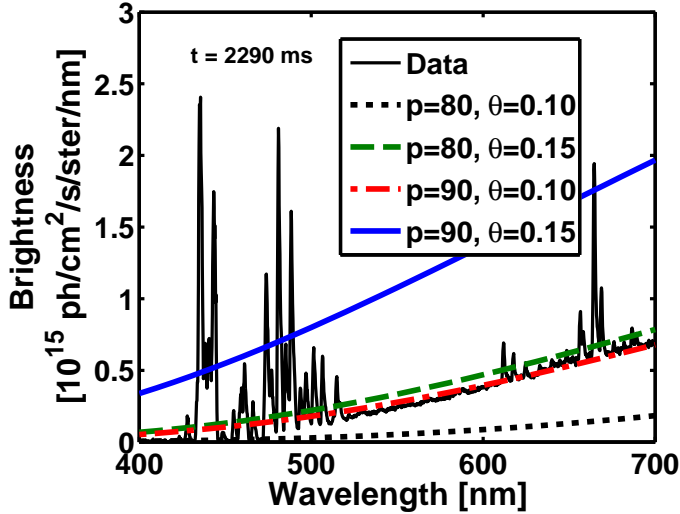}
\caption{(Color online) Measured visible spectrum in DIII-D during the runaway plateau at t=2290 ms in shot 146704. Spectra from several mono-energetic populations calculated using \PankTwo are also shown. The number of runaways used to obtain the spectra was 1\% of $n_r$ calculated from the runaway current (assuming $r_{re}=20\,$cm). \label{fig:diiid_single}}
\end{figure}

For comparison, we also fit the experimental data with synchrotron
spectra from a mono-energetic runaway population (using
Eq.~\ref{eq:brightness}), for different particle energies and
pitch-angles. As in Ref.~\cite{yu} we assume that 1\% of the runaway
population (calculated with $r_{re}=20 \;\rm cm$) has the specific
energy considered. The results are shown in
Fig.~\ref{fig:diiid_single}. This fitting procedure gives a lower
estimate for the maximum runaway energy, at about 40-50 MeV, depending on
pitch-angle.

\subsection{Effect of wave-particle interaction}
\label{sec:wave}

\begin{figure}
\includegraphics[width=0.47\textwidth]{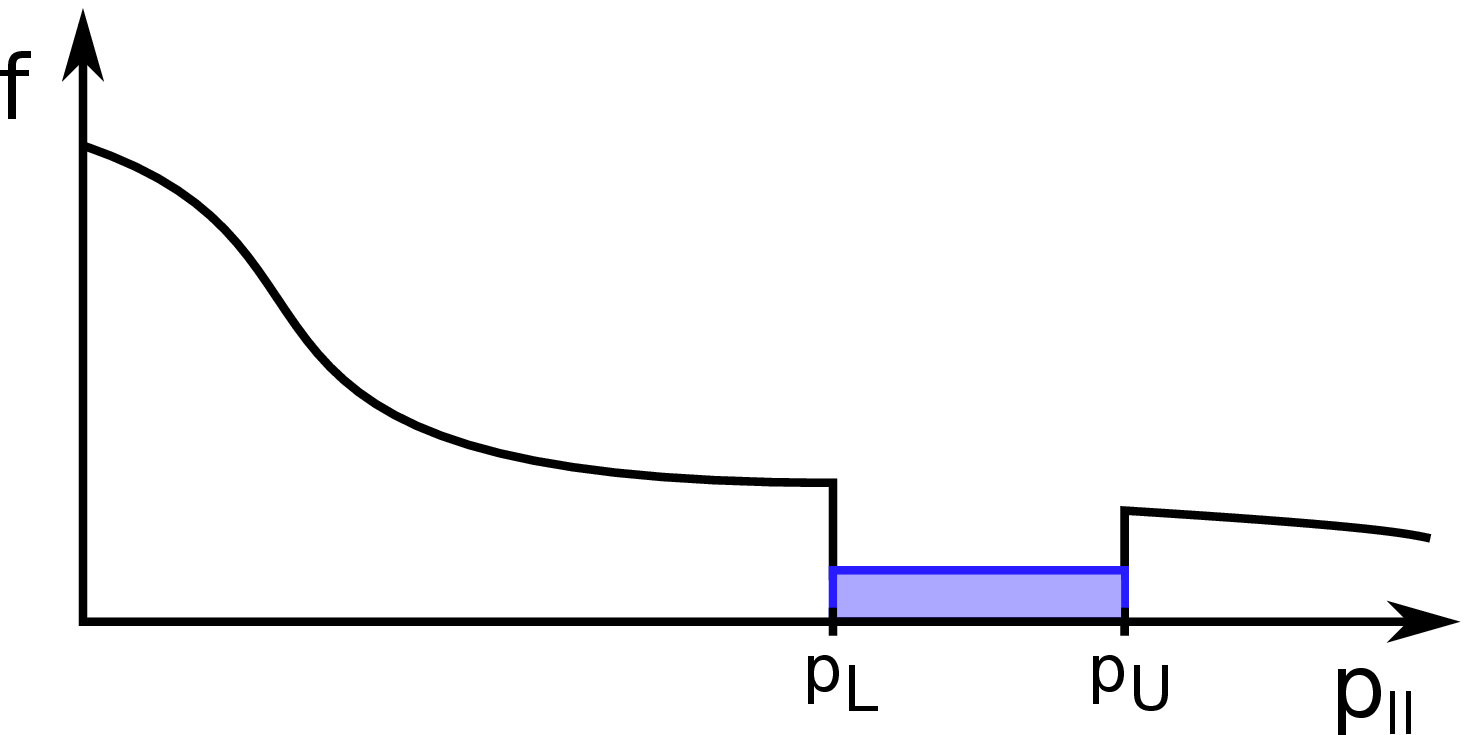}
\includegraphics[width=0.47\textwidth]{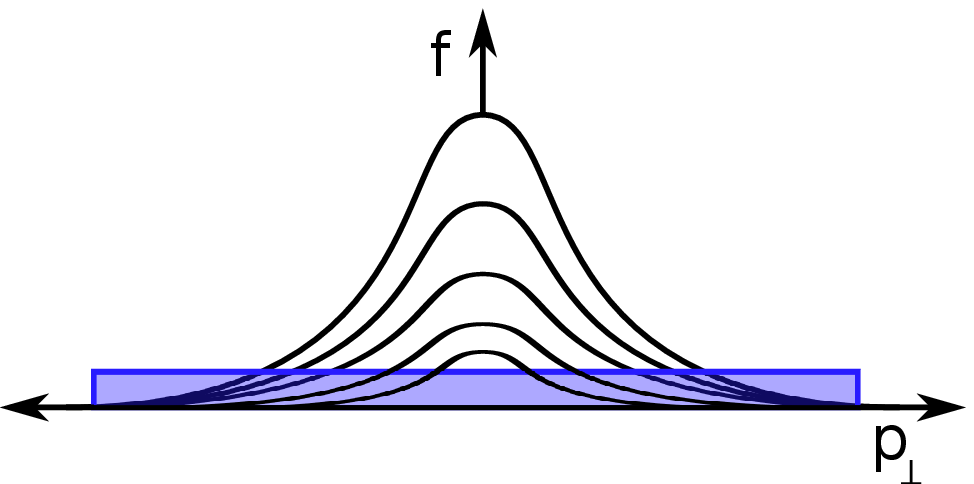}
\caption{(Color online) Schematic runaway distribution with modifications emulating the effects of wave-particle interaction.  \label{fig:Modified_distribution}}
\end{figure}

Another instance where the synchrotron spectrum from a complete
runaway distribution is useful is in investigations of mechanisms that
affect the shape of the distribution itself. One such mechanism is
resonant wave-particle interactions, and here we consider their effect
on the synchrotron spectrum through a modification of part of the
distribution given in Eq.~(\ref{eq:anal_ava}). A runaway distribution
is normally strongly peaked around the parallel direction ($\chi=1$),
i.e. it has a high degree of anisotropy in momentum space (see for
instance Fig.~\ref{fig:contours}). Wave-particle interaction tends to
drive the distribution towards isotropy through pitch-angle scattering
of electrons with resonant momenta \cite{pokol}. A simple way to
simulate the decrease in anisotropy is to introduce a flat profile in
part of momentum space, as indicated in
Fig.~\ref{fig:Modified_distribution}.

The usual integral for the total emitted power,
Eq.~\eqref{eq:emission}, is split up into three regions in
momentum-space. The first and third parts remain unmodified, with the usual
distribution function $f_{RE}$. In the second (middle) part,
the distribution function is assumed to be flat. We denote the lower
and upper boundaries of this region $p_{L}$ and $p_{U}$, respectively.
The momentum space volume of the shaded block in the figure should be
the same as that of the part of the distribution it replaces, which
gives us a condition from which to calculate the appropriate height of
the block. The integration of the normal distribution is taken over
the entire $\chi$-range ($\chi \in [0,1]$). As the distribution
decreases exponentially with decreasing $\chi$, the contribution from
particles with low $\chi$ is very small. When the modifications are introduced, however, the contribution
could be substantial, and we need to restrict the extent of
the block for the modified part of the distribution in $\chi$ ($\chi \in
[\chi_\text{min},1]$). The introduction of
$\chi_{min}$ can be seen as a compensation for the fact that in reality
the pitch-angle scattered particles are not evenly distributed in
$\chi$.  Letting $f_{c}(p,\chi)=h$ be a constant distribution where $h$
represents the height of the block, and equating the momentum space
volume of the block with that of the part of the distribution it
replaces, we have
\begin{align}
V & =2\pi\int_{0}^{1}\!\!\!\int_{p_{L}}^{p_{U}}\! f(p,\chi)\, p^{2}\mbox{d}p\mbox{d}\chi=2\pi\int_{\chi_\text{min}}^{1}\!\int_{p_{L}}^{p_{U}}\! f_{c}(p,\chi)\, p^{2}\mbox{d}p\mbox{d}\chi
& =h\cdot\frac{2\pi}{3}(1-\chi_\text{min})(p_{U}^{3}-p_{L}^{3})\,.\label{eq:Volume}
\end{align}
We may solve this for $h$, and obtain
\begin{equation}
h=\frac{3\int_{0}^{1}\int_{p_{L}}^{p_{U}}f(p,\chi)p^{2}\mbox{d}p\mbox{d}\chi}{(1-\chi_\text{min})(p_{U}^{3}-p_{L}^{3})}\ 
\end{equation}
as the block height that conserves the total number of particles. We
emphasize that the above modification represents a ``worst case
scenario'' in terms of the effect on the spectrum. In a more realistic
case, the modifications would be less severe.

\begin{figure}
\includegraphics[width=0.55\textwidth]{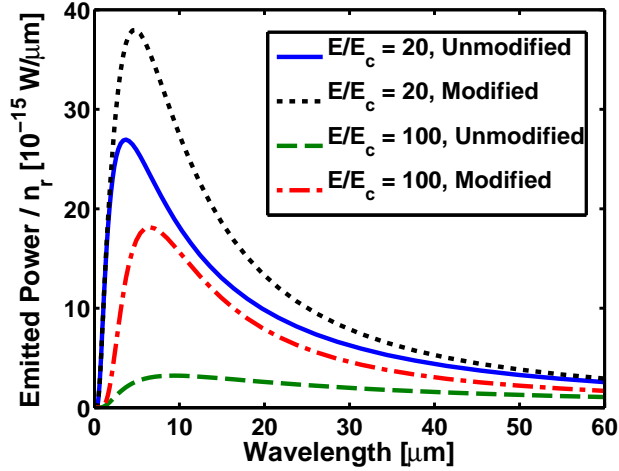}
\caption{(Color online) Synchrotron spectra from unmodified and modified runaway
  distributions for different electric field strengths. The parameters
  used where $\pmax=50$, $p_L=25$, $p_U=35$, $\zeff=1.6$, $n_{e}=3\power{20}\,$m$^{-3}$,
  $T=10\,$eV, and $B=3\,$T. For these parameters the critical field is
  $E_{c}=0.15\,$V/m. The maximum pitch-angle for the particles in the
  modified region was set to $\pp/\ppa=0.2$.}
\label{mod_vs_normal}

\end{figure}

The analytical avalanche distribution (Eq.~\ref{eq:anal_ava}) was modified according to the
above, with $p_{L}=25$ and $p_{U}=35$ since this is a typical range
where wave-particle interactions manifest \cite{pokol}. The maximum
pitch-angle in the modified region was set to $\pp/\ppa=0.2$
($\chi_\text{min}=0.98$), which is qualitatively consistent with
experimental estimates of the maximum runaway pitch-angle
\cite{jaspers,yu}.  In Fig.~\ref{mod_vs_normal}, modified
distribution-integrated synchrotron spectra are shown and compared to
those of unmodified distributions. From the figure it is clear that there is an appreciable
increase in the average emission of the runaways as a result of the
modifications to the distribution. Again, this increase is related to the
pitch-angle dependence of \Bekefi. The isotropization
broadens the distribution in pitch-angle which leads to a higher
average emission. Due to the difference in the synchrotron
spectrum, the onset of a particle-wave resonance should be
detectable. However, as we have seen before, there are also other
changes in plasma parameters that could have a similar effect on
the synchrotron emission.

Our goal in this exercise is not to explore the
parameter space of artificially modified distributions - the
modifications introduced above are too crude to lead to
quantitative conclusions - but rather to illustrate the sensitivity of
the synchrotron spectrum to the details of the runaway
distribution. The analysis here shows that the spectrum from a distribution
modified by particle-wave interaction can imply runaway
parameters distinctly different from those that are actually present,
especially if only a limited part of the spectrum is
considered. Failure to include such effects can thus lead to incorrect
conclusions regarding the runaway beam properties.

\section{Conclusions}
\label{sec:conclusions}
The synchrotron emission spectrum can be an important diagnostic of
the runaway electron population. In some previous work, synchrotron
spectra have been interpreted under the assumption that all runaways
have the same energy and pitch-angle.
In practice, however, runaway electrons have a wide distribution of
energies and pitch-angles.  When taking into account the full
distribution, the most suitable approximative emission formula may not
be the one that has been used in previous work
(\PankOne). Instead, depending on the major radius of the
device and the actual runaway electron distribution, either
\Bekefi (for large devices) or \PankTwo
(for medium-sized devices) are more suitable.  Although the single
particle synchrotron emission formulas do not depend on the plasma
temperature, effective charge, density or electric field strength, the
total synchrotron emission is sensitive to these parameters, as they determine the shape of the runaway distribution.

We have shown that the single-particle emission overestimates the
synchrotron emission per particle by orders of magnitude, and the
wavelength of the peak emission is shifted to shorter wavelenths
compared to the spectrum from an avalanching runaway electron
distribution. We have also illustrated that using the slope of the
spectrum for estimating the runaway energy can be misleading, and in
general one should calculate the emission from an assumed
approximative distribution and iteratively find the maximum runaway
energy to fit the synchrotron spectrum. Finally, through a comparison
with an experimental synchrotron spectrum from DIII-D, we have
estimated the maximum runaway electron energy in that particular
experimental scenario to be around 65 MeV.

\section*{Acknowledgments}
The authors are grateful to Y.~Kazakov for fruitful discussions.
This work was funded by the European Communities under Association
Contract between EURATOM, HAS  and {\em Vetenskapsr{\aa}det}. The views and
opinions expressed herein do not necessarily reflect those of the
European Commission.  M.L. was supported by the United States
Department of Energy's Fusion Energy Postdoctoral Research Program
administered by the Oak Ridge Institute for Science and Education. E.H. was supported in part by the United States Department of Energy under DE-FG02-07ER54917.

\bibliographystyle{unsrt}

\end{document}